\documentclass[useAMS,usenatbib]{mn2e}

\usepackage{graphicx,longtable}
\usepackage{natbib}

\newcommand{\km}{\,\mbox{km}\,\mbox{s}^{-1}}
\def\Ha{H$\alpha$}

\title[A new catalogue of polar-ring galaxies]{A new catalogue of polar-ring galaxies selected from the SDSS.
\thanks{Partly based on observations collected with the 6m  telescope of the Special Astrophysical Observatory of the  Russian Academy of Sciences which is operated under the financial support of Science Department of Russia (registration number 01-43)}}
\author[Moiseev et al.]
{Alexei V. Moiseev, $^1$\thanks{moisav@gmail.com} Ksenia I.
Smirnova,$^2$
 Aleksandrina A. Smirnova,$^{1}$
 \newauthor and Vladimir P.  Reshetnikov,$^{3,4}$  \\
$^1$Special Astrophysical Observatory,  369167 Nizhnii Arkhyz,  Karachaevo-Cherkesskaya Republic,  Russia\\
$^2$Ural State University, Lenin ave. 51, 620083 Ekaterinburg, Russia\\
$^3$St. Petersburg State University, Universitetskii pr. 28,  198504 St. Petersburg, Stary Peterhof,  Russia\\
and Isaac Newton Institute of Chile, St.Petersburg Branch\\
$^4$ Observatoire de Paris, LERMA, CNRS, 61, av. de l'Observatoire, Paris, F-75014, France
}
\begin{document}

\date{Accepted 9 Jule 2011  Received 21 Apr 2011}

\pagerange{\pageref{firstpage}--\pageref{lastpage}} \pubyear{2011}

\maketitle

\label{firstpage}

\begin{abstract}
Galaxies with polar rings (PRGs) are a unique class of extragalactic
objects allowing to investigate a wide range of problems, linked
with the formation and evolution of galaxies, and to study the
properties of their dark haloes. The progress in the study of PRGs
is constrained by a small number of known objects of this type.
Whitmore et al. (1990) compiled a catalogue and photographic
atlas of PRGs and related objects that included 157 galaxies.
Up to date, we can only attribute about two dozens of
kinematically-confirmed galaxies to this class, mostly from this catalogue.

We present a new catalogue of PRGs, supplementing the catalogue of
Whitmore et al. (1990), and significantly increasing the number of
known candidate PRGs. The catalogue is based on the results of the
 original Galaxy Zoo project, under which the volunteers performed a visual
classification of nearly a million galaxies from the SDSS. Based
on the preliminary classification of the Galaxy Zoo, we viewed
more than 40\,000 images of the Sloan Digital Sky Survey (SDSS)
and selected  275 galaxies, included in our catalogue.

Our Sloan-based Polar Ring Catalog (SPRC) contains  70 galaxies that we classified as
``the best candidates'', among which we expect to have a very high
proportion of true PRGs, and  115 good  PRG candidates.   53
galaxies are classified as PRG related  objects
(mostly, the galaxies with strongly warped discs, and mergers). In
addition, we identified  37 galaxies that have their presumed polar
rings strongly inclined to the line of sight (seen
almost face-on).  The SPRC objects are on the average fainter and
located further away than the galaxies from the catalog by
Whitmore et al. (1990), although our catalogue does include
dozens of new nearby candidate PRGs.

The new catalogue significantly  increases the number of genuine
PRG candidates, and may serve as a good basis both for the further
detailed study of individual galaxies, and for the statistical
analysis of PRGs as a separate class of objects.

We performed spectroscopic observations of six galaxies from the
SPRC at the 6-m BTA telescope. The existence of polar rings was
confirmed in five galaxies, and one object appeared to be a
projection of a pair of galaxies. Adding the literature data, we
can already classify 10 galaxies from our catalogue to the
kinematically-confirmed PRGs.

\end{abstract}


\section{INTRODUCTION}

Galaxies with polar rings (PRGs) are an interesting case of
peculiar systems that reveal outer rings or discs of gas, dust and
stars, rotating in the plane approximately perpendicular to the
disc of the main (or host) galaxy. It is believed that the
formation of PRGs is in most cases caused by galaxy mergers with
the corresponding direction of angular momentum, the accretion by
the host galaxy of the companion's matter, or gas filaments from
the intergalactic medium  \citep[see references in the review by ][]{Combes06}.
It is shown
that in the case of an oblate or triaxial gravitational potential,
stable orbits exist in the polar plane, hence, the captured
material of the companion will be rotating here long enough. If
the orbital plane is notably different from polar, then the formed
ring would relatively quickly precess to the Galactic plane
\citep{Steiman1982}.

A detailed examination of the structure and dynamics of PRGs both
clarifies the role of interactions and mergers in the current
evolution of galaxies, and also allows to study the features of
the distribution of gravitational potential at large radii. Since
we are observing here a circular rotation in two mutually
perpendicular planes, it becomes possible to study the
three-dimensional distribution of mass in the galaxy. Particularly
we can determine the shape of the dark halo: its oblateness,
elongation, and deviations from axial symmetry
\citep{SackettPogge1995,Iodice03,Combes06}.

In addition, there are
recent indications that some relatively massive polar rings are
formed by a cold accretion of gas from the filaments of the intergalactic medium.
In this scenario, \citet*{Maccio2006} and \citet{Brook2008} were able to construct
an object very similar to the well-known PRG NGC~4650A using
cosmological hydrodynamic simulations. Recently, additional
observational arguments supporting the model of accretion from gas
filaments were presented.  For instance, \citet{Spavone2010}
suggested that NGC~4650A has the gas metallicity lower than a spiral galaxy disc of the same total luminosity.
Also, \citet{Stanonik2009} have found in the galaxy SDSS J102819+623502
a polar disc of neutral hydrogen aligned nearly perpendicularly to a cosmological wall situated
between two voids. Therefore, a further study of PRGs will help to better perceive the causes
(and the source) of the accretion of gas from the intergalactic
medium, and will help to resolve some cosmological puzzles
of galactic evolution \citep*{Combes2008}.

Although the first credible evidences of the existence of polar
rings were obtained quite long ago \citep{SchechterGunn1978},
their statistical study began only after \citet{Whitmore1990} have
published a catalogue of PRG candidates -- the Polar Rings
Catalog=PRC. In total, their catalogue contained 157 objects, from
which only six (PRC A category) already had a kinematic confirmation
(rotation, detected in two orthogonal planes), and 27 galaxies were
included the category of `good candidates' (PRC B). The greater
part of the list was occupied by 73 ``possible candidates'' (PRC
C) and 51 ``related objects'' (PRC D). Over the following twenty
years, several groups were involved in the study of kinematics of
catalogue objects (mostly -- in the northern celestial hemisphere)
both using the methods of optical spectroscopy and radio
interferometry. However, the number of kinematically confirmed
PRGs is still meager. In particular, the outer polar rings were
confirmed in about 20 galaxies from the PRC \citep*[]{Resh2011},
including several ``related systems'', where the polar structures
are only forming. In addition, six other galaxies contain inner
polar discs with radii smaller than $1-3$ kpc.

There are even fewer PRGs studied in detail, we can currently
speak here of less than a dozen galaxies only. The question of the
shape of their dark halo remains open. On the one hand, the
statistical analysis of the distribution of integral parameters of
PRGs in the Tully-Fisher diagram indicates that the outer halo is flattened to the plane of the
polar ring \citep{Iodice03,Resh2004}.
This means that the major principal plane of the halo is
orthogonal to the disc of the central galaxy, which in itself is
unusual. On the other hand, the results of a study of several
galaxies with strongly warped HI discs, such as NGC~2685 = PRC A-3
\citep{Jozsa2009}, NGC~3718=PRC D-18 \citep{Sparke2009}, and
NGC~4753=PRC D-23\citep*{Steiman1992} indicate a more or less
spherical halo.  The complexity of the problem and the need for a
detailed analysis of each galaxy is illustrated by the case of NGC4650A.
\citet*{Whitmore1987} have earlier noted that the
gravitational potential of the galaxy is almost spherical, but the
analysis of  kinematics of more extended structures indicates a
very strong oblateness of the halo, oriented in accordance with
the polar ring \citep{Iodice2010}.

The gravitational stability of polar rings is a subject of debate
as well. \citet{Wakamatsu1993} shows that while the
polar ring gas clouds pass through the gravitational well of the stellar disc, the
gas generates shock waves that ``clean out'' the inner region of
the ring. Obviously, they can additionally induce star formation
in the ring. Numerical hydrodynamic models by \citet*{Theis2006}
also indicate the possibility of formation of a stable spiral
structure  in the polar ring of   NGC4650A. Unfortunately, the observational
evidence for the existence of spirals in the polar rings is yet
rare and contradictory, since to confirm this hypothesis, the
polar disc has to be sufficiently inclined to the line of sight.
In addition, there are no reliable observational estimates of the Toomre's
parameter $Q$ for the polar rings and discs as yet.

There are PRGs, discovered outside the catalogue by
\citet{Whitmore1990}.  This is a dwarf galaxy NGC~6822 of the
Local Group, in which the outer HI disc rotates in the plane,
polar to the stellar disc \citep{Demers2006}.   A similar but larger structure
 is detected around a more distant galaxy SDSS J102819.24+623502.6
\citep{Stanonik2009}. Ultraviolet images of  NGC~4262 reveal signatures
of star formation in the gas ring surrounding it  \citep{Bettoni2010}.
The most  distant confirmed published  polar ring ($z=0.06$) was studied by
\citep{Brosch2010}.  In the Hubble deep fields three even more distant
PRG candidates were discovered at  $z=0.6-1.3$, which are still
awaiting confirmation \citep{Resh1997,Resh2007}.

Two directions can be envisioned in the prospects for the  further
studies of PRGs. First of all, a detailed study of the known
candidates and their environments
needs to be done using the data on the morphology and
kinematics in different spectral ranges, from the UV to radio.
Secondly, we have to expand the list of candidates, to both
clarify the luminosity function of PRGs, and to move towards
higher redshifts, but as well to search for objects in which both
the ring and the galaxy are ``conveniently'' oriented to the line
of sight, allowing to simultaneously study their kinematics and
structural details. The PRC was
based on the study of photographs of individual galaxies. In the
modern era, it is reasonable to use digital sky surveys, such as
the SDSS for these purposes.

Note that we do not
 discuss in this paper the so-called inner polar rings and
discs, observed in the circumnuclear regions of nearby, typically
early-type galaxies \citep[see references and discussion in] []
{Corsini2003, SilAfan2004}. Such structures \citep[$\sim30$ of which are
already known, see][]{Moiseev2010} are lost in the bright
background of the bulge, and are discovered mainly due to their
kinematics, decoupled from the  galaxy disc.

In this paper, we present a new list of PRG candidates, a few of which have already been confirmed.
Section  \ref{sec_catalog} describes the technique of
catalogue compilation using the data from the  Galaxy Zoo project.
Section \ref{sec_ABCD} describes the division of catalogue objects
into several types.  In Section \ref{sect_obs} we  present the information
about five galaxies for which  there already exist detailed studies of the internal
kinematics. We managed as well to perform spectral observations of six
galaxies with the 6-m BTA telescope of the SAO RAS. Five of the observed objects were confirmed to be
classical PRGs, and one turned out to be a projection of an
interacting pair of galaxies. Section \ref{sec_conclusion} briefly discusses the results of this paper.

\begin{figure}
\includegraphics[width=0.45\textwidth]{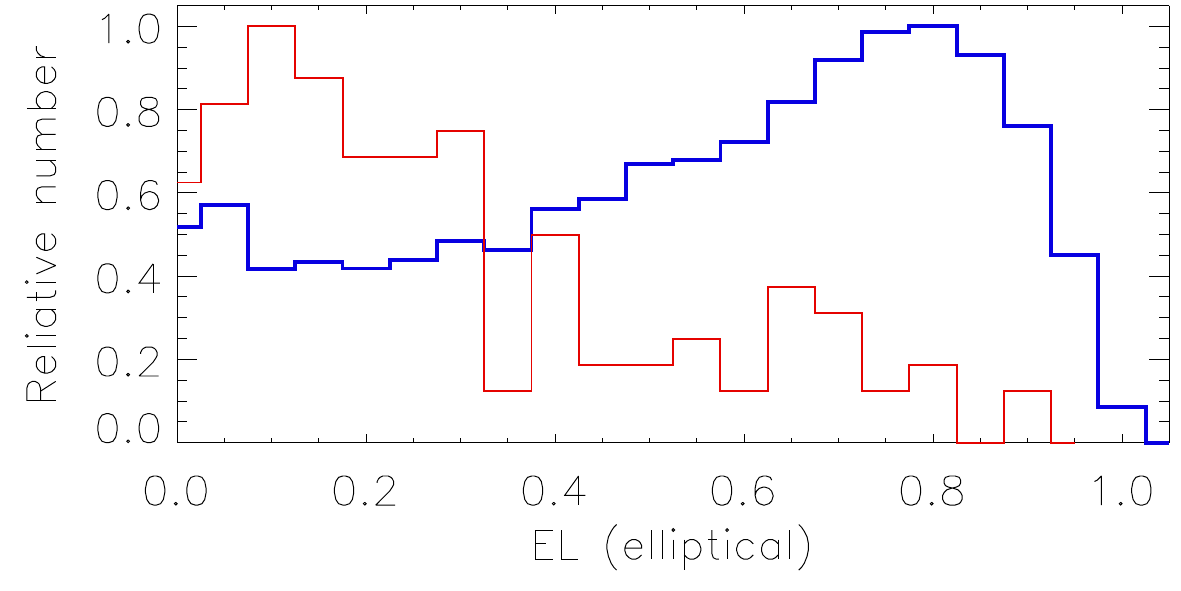}
\includegraphics[width=0.45\textwidth]{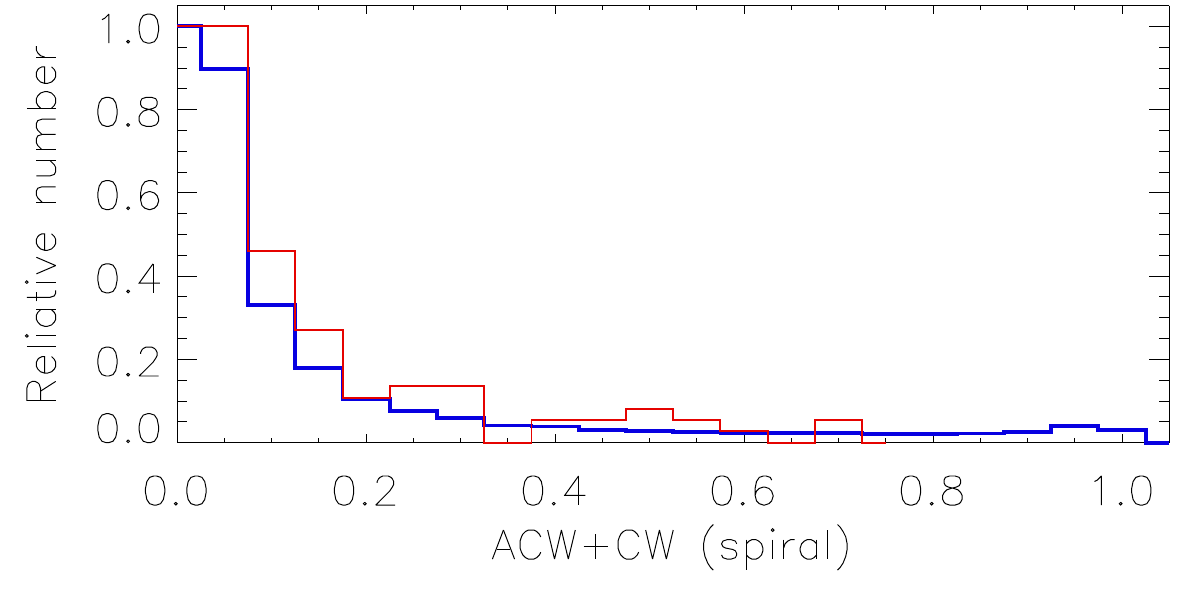}
\includegraphics[width=0.45\textwidth]{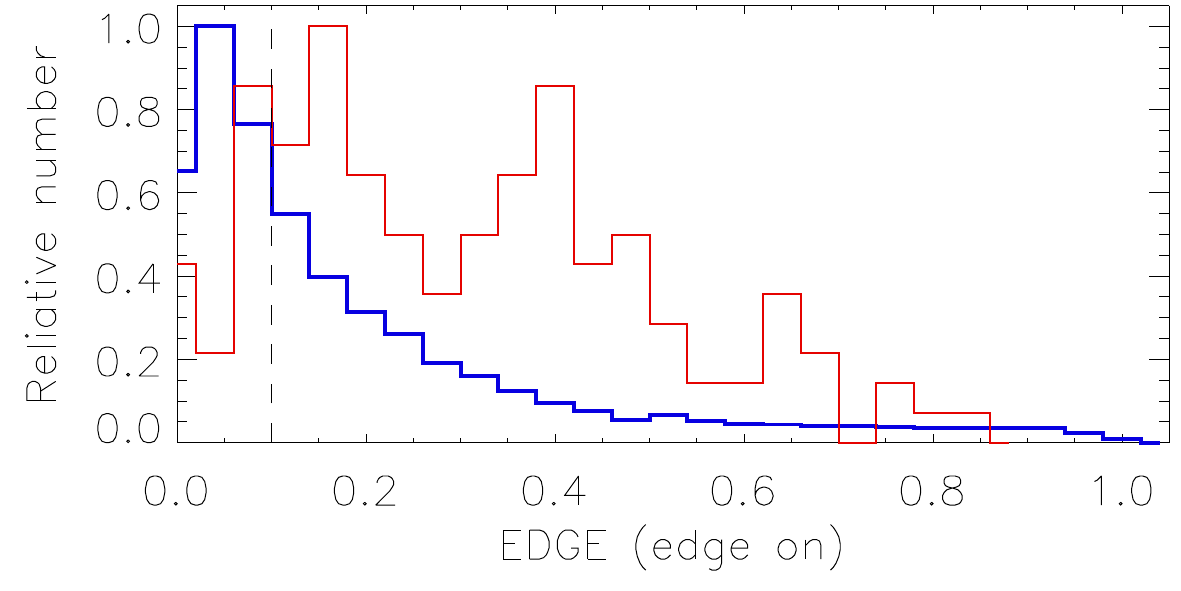}
\includegraphics[width=0.45\textwidth]{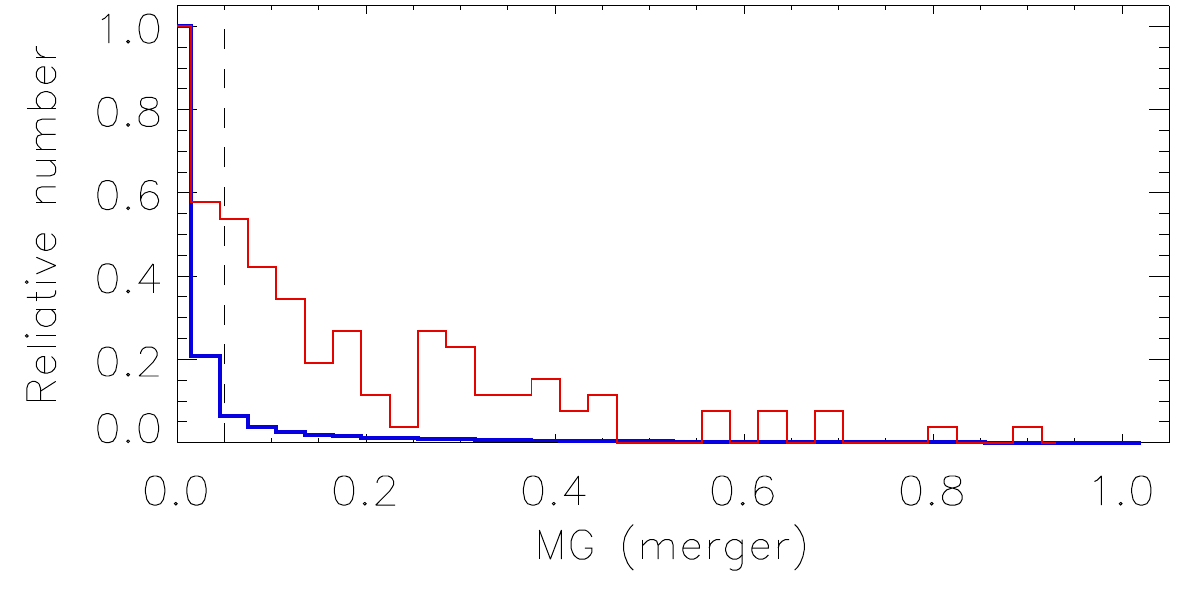}
\includegraphics[width=0.45\textwidth]{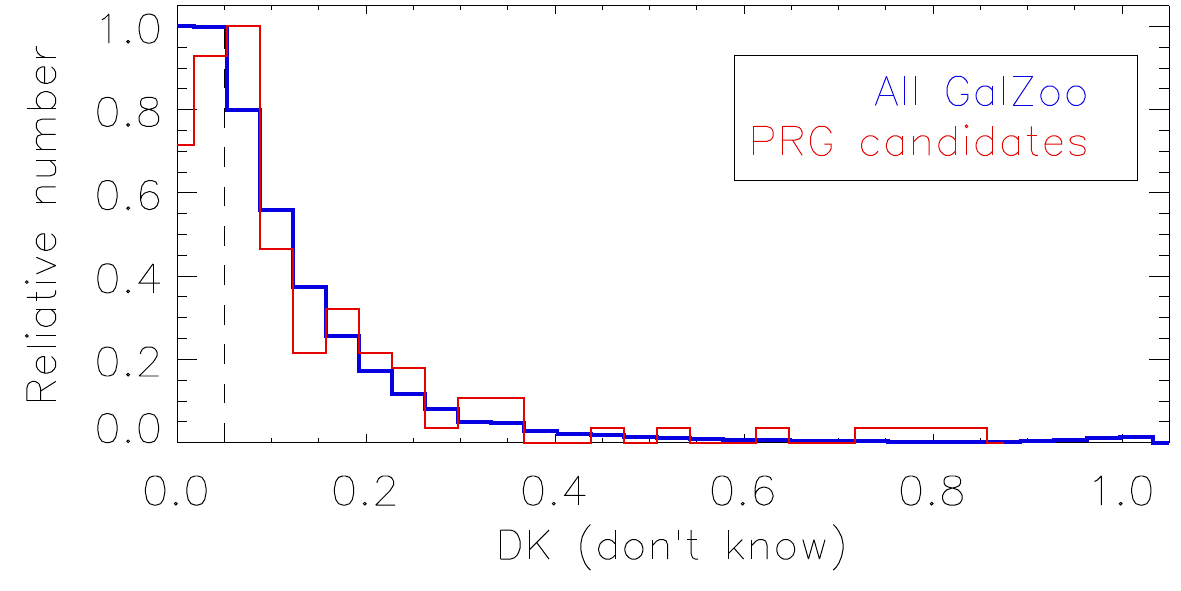}
\caption{Histograms of distribution of relative galaxy numbers within each
type. The blue line -- for all the  893,212 galaxies of the Galaxy
Zoo, red -- for 126 PRG candidates from the reference sample. Dashed
lines show the selection criteria.} \label{fig_zoo}
\end{figure}

\section{Catalogue Compilation}
\label{sec_catalog}

\subsection{Preliminary search}
\label{sect_preliminary}

The SDSS covers a significant part of the
celestial hemisphere ($\sim1/4$ for Data Release 7) and contains
optical images of millions of galaxies \citep{sdss7}.
Unfortunately, there are no sufficiently reliable methods for
accurate automatic classification of galaxy images by
morphological type. Although such algorithms are developed, their
reliability  is not yet sufficient for mass use  \citep[see
discussion and references in][]{GalZoo2011}. Moreover, the images
of such peculiar objects as PRGs are very complex, in many cases
an analysis of features of low surface brightness is required
to attribute a given galaxy to the PRG candidates.

Fortunately, we now have an opportunity to use the results of the
unprecedented Galaxy Zoo project\footnote{\texttt{"http://www.galaxyzoo.org/}}, in which hundreds of thousands of
volunteers around the world are visually classifying the SDSS
galaxies. Of course, they are not engaged in a separate search for
galaxies with polar rings, but they do submit many expressive
examples to the Internet-forum dedicated to the ring galaxies. Most of the
galaxies listed are collisional rings, or rings on the bar
resonances. Note that some galaxies were already mentioned by the forum as the
possible PRG candidates, including Hoag-type galaxies. But looking at hundreds of these images, we were able
to select 92 candidate PRGs that were not included in the PRC
catalogue. We were guided by the following selection criteria:

\begin{enumerate}
\item The presence in the image of two almost orthogonal discs,
taking into account their inclination to the line of sight. The
influence of the projection effects is  illustrated quite well in
Fig.~1 from \citet{Whitmore1990}.
\item The photometric centres of both subsystems coincide with an accuracy of about 1--3 arcsec.

\item The apparent diameter of the galaxy is at least 10 arcsec
(otherwise it is difficult to examine the details of its structure).
\end{enumerate}

 To the final list we have also added the
candidates available in the SDSS, but discovered in previous studies:
three galaxies, found by examining the images of peculiar objects and
galaxies with outer rings from the lists by
\citet{NairAbraham2010}, who visually classified  $14\,034$
galaxies in the SDSS/DR4; three candidates that we found analyzing the
notes to the MCG catalogue \citep{MCG};  SDSS
J132533.22+272246.7 found by  \citet{Finkelman2011} viewing the SUBARU deep field; another galaxy from the
ESO Press Release 14/98  -- SDSS J000911.57-003654.7 \citep{Resh2011}; a distant PRG SDSS
J075234.33+292049.8  \citep{Brosch2010}, mentioned in the Introduction  and also 14 galaxies kindly provided us by Ido Finkelman \citep[some of his galaxies are studied in]{FinkelmanNEW}.

As a result, we compiled a list of 122 galaxies, which in itself
significantly expands the list of ``genuine'' candidates from the
PRC, since the majority of discovered objects  corresponds to the
PRC-B and PRC-C categories. However, this list obviously suffers
from a heavy incompleteness, since only a small fraction of the
Galaxy~Zoo volunteers reports on unusual galaxies in the
forum.

\subsection{Analysis of the Galaxy Zoo sample}

\label{sect_zoo}

\citet{GalZoo2011} have recently reported the results of a simple
morphological classification for almost 900,000 galaxies, made by
Galaxy Zoo volunteers. They have published lists of 667,945
galaxies with known spectral redshifts  (in the range $0.001<z<0.025$) and 225,268 galaxies
without the spectra from the SDSS/DR7. The catalogue contains  the majority  of
extended objects, having in the $r$-band Petrosian magnitudes
brighter than  $17.77^m$, as well as the
galaxies from the SDSS spectral survey. The classification of
galaxies was carried out by the following simple types: $EL$ --
elliptical; $CW/ACW$ -- clockwise and anticlockwise spirals;
$EDGE$ -- spiral, seen edge-on; $MG$ -- interacting
systems; $DK$ (don't know) -- unidentified. At a first glance it
seems that this classification is too coarse for our purposes, but
the fact that for each galaxy a large number (as a rule -- a few
dozen) of independent evaluations were carried out greatly
simplifies the situation.  The objects are hence characterised not
simply by one of the above types, but rather by a fraction of
votes cast for each type, which can be called, subject to
restrictions, the probability of belonging to a particular type.

Fig.~\ref{fig_zoo} demonstrates these probabilities for all the
Galaxy Zoo galaxies and for the ``reference sample'' consisting of
the known PRG candidates, classified in the Galaxy Zoo. The
reference sample contains 126 objects and includes galaxies,
selected in Section 2.1 (for 103 of them the Galaxy Zoo has a
classification), and all the galaxies from the PRC catalogue,
classified in the Galaxy Zoo\footnote{From 157 PRC galaxies, there
exist SDSS/DR7 images only for 27, out of which 23 were classified
in the Galaxy Zoo.}. For both samples, these distributions are
significantly different, except for the probability of belonging
to spiral galaxies (CW+ACW). Peculiar shapes of PRG candidates do
not allow the majority of respondents to rank them as elliptical
galaxies, hence for them $EL<0.85$. For the same reason, PRG
candidates reveal a much broader distribution by the probability
of merging, while most of the Galaxy Zoo galaxies are concentrated
towards $MG\approx0$. Among the PRGs there are almost no spirals,
oriented face-on (i.e., for which $EDGE\approx0$), so the
corresponding distribution is not only wider, but its peak is
notably shifted towards the higher $EDGE$ values  because the edge-on configuration is much easier for the detection of polar structures. The
distributions by the $DK$ parameter do not vary as much, but we can
still see that the peak of the PRG candidate distribution is
shifted from zero to $DK\approx0.07$. We can explain this by the
fact that among the dozen of volunteers, classifying each PRG, at
least 1--2 respondents marked its unusual shape by belonging to
the $DK$ type.

As a result, we can formulate the criteria for the selection of
galaxies, similar (within the presented types) to the already
known candidate PRGs, and hence greatly reduce the number of
images for further examination. We adopted   the following
criteria for the selection of peculiar galaxies:

\begin{equation}
EDGE\ge0.1,
MG\ge0.05,
DK\ge0.05
\label{e}
\end{equation}

We introduced no restrictions on the type $EL$ membership, since it
follows from (\ref{e})  that $EL<0.8$. In the full Galaxy Zoo
sample, 30,084 galaxies with known redshifts, and 11,874 objects
without spectral data satisfy the (\ref{e})  criteria.  From 126 PRG
candidates, shown in Fig.~\ref{fig_zoo} only 47 galaxies, i.e.
37 per cent  of the sample satisfy these criteria. Thus, if we find ALL
the potential PRGs from the selected 41,958 galaxies, their actual
number in the whole Galaxy Zoo catalogue will be underestimated
 $1/0.41\approx2.7$ times. But we have to agree to this, since the
use of less stringent criteria greatly increases the number of
galaxies to examine. In this way, if we reduce all the criteria in
(\ref{e}) by a mere 0.015, they will be met by 68,412 objects in
the total sample, while the number of objects in the PRG reference
sample, falling under the criteria will increase only up to 48 per cent.

We viewed the images of all the galaxies, selected in agreement
with (\ref{e}). The web interface SDSS Image Tool we used allows
to display on one page 25 colour images, composed of the images in
the $g,r,i$  filters at once. Looking through the images,
we selected candidate PRGs mainly guided by the selection criteria, listed
in Section 2.1, as well as by a general similarity with the PRC
catalogue objects. After examining nearly 42,000 images, we
selected about 400, which were further investigated in more detail
for the final selection.

\begin{figure}
\includegraphics[width=0.45\textwidth]{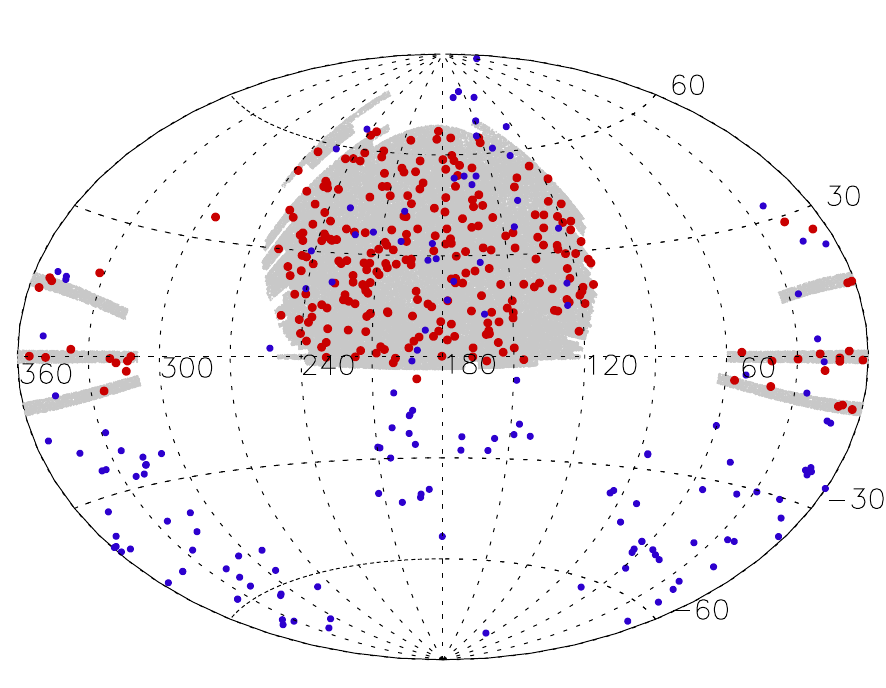}
\caption{Apparent distribution of PRG candidates in the sky in
equatorial coordinates. The gray shaded area marks the Galaxy Zoo
region. Blue dots are the objects from \citet{Whitmore1990}, red
dots are the galaxies from the new catalogue.} \label{fig_sky}
\end{figure}

\section{Catalogue Description}
\label{sec_ABCD}

The final list of PRG candidates contains  275 galaxies, including
those  selected in Section 2.1. Thus, we have been able
to triple the original list, which was largely based on what was
found at the Galaxy Zoo forum of ring galaxies.
Fig.~\ref{fig_sky} shows the distribution of catalogue objects
on the celestial sphere. We can see that they uniformly fill the
region, covered by the Galaxy Zoo, except for a small fragment
around ($\alpha=140^\circ,\,\delta=7^\circ$) close to the Galactic
plane\footnote{ Several point outside the Galaxy Zoo region in
Fig.~\ref{fig_sky} belong to a galaxy, found by the Galaxy Zoo
volunteers in the  SDSS/SEGUE data and also to 6 galaxies from the SDSS/DR8
 presented on the Galaxy Zoo forums.}.

By analogy with the PRC, we named our list SPRC = Sloan-Based
Polar Rings Catalogue.  The catalogue is divided into four unequal
groups (best candidates, good candidates, related objects and
possible face-on rings). Since the division into these types is often
ambiguous, in contrast to \citet{Whitmore1990} we did not assign
each group with its own index, but used a continuous numeration of
objects instead. However, for convenience sake, the objects in
each group are ordered by their R.A. The lists of galaxies themselves are
given in Table~\ref{tab_ABCD}, where in addition to numbers in our
catalogue and equatorial coordinates, we give the total apparent
magnitude in the $r$ filter (corrected for the Galaxy extinction,  as listed in the
SDSS/DR8 \texttt{galaxy} table)\footnote{ Magnitudes for SPRC-85 and SPRC-235 are in $B$
filters  (NED), because for these galaxies with a  very complex morphology the SDSS catalogue
provides model fluxes lower than 20~mag, which seems unrealistic.},
redshift and the name of the galaxy from well-known catalogues (NGC, UGC, CGCG,
etc). Redshifts are given according to the NED database. If the
NED data are missing, we use the SDSS data. For several galaxies
no SDSS spectra exist.

Fig.~\ref{fig_example} demonstrates some of the most typical
examples of different types of catalogue objects. Images of all
the catalogue galaxies are given in Fig.~\ref{fig_atlas}
(presented in the electronic version of the paper). Combined
colour JPEG images are provided by   the SDSS/DR8, which was released in the early 2011
\citep{sdssdr8}, and only in a few cases of mosaic creation
problems the data were taken from the SDSS/DR7.

\subsection{Division into types}

The division of candidates into types has a significant
uncertainty, since it is based on the appearance of galaxies in
the optical images, not employing explicit numerical criteria.
Taken that, we tried not to discard any ``suspicious'' objects,
which seemed to be related to PRGs. Nevertheless, like in the old
PRC catalogue, the introduction of several categories is useful in
terms of selecting objects for a further detailed study. Keep in
mind that only a detailed analysis of internal kinematics can
confidently confirm the presence of decoupled
components in a galaxy. For some galaxies, such a confirmation is
already available.

In the cases when there was additional data provided on individual
galaxies in the publications,   we have used this information to refine the classification.
Namely, we classified the galaxy SPRC-1 in which
the polar component is vaguely expressed (the dust lane is mostly visible)
to the group of best candidates, as \citet{Resh2011}  have shown
the presence of an extended ring in the optical images (see as well the ESO Press Release 14/98). The
belonging of the SPRC-7 galaxy to PRGs was proved by spectral
observations by \citet{Brosch2010}, hence we did not classify it
into a group of candidates with nearly face-on rings. The polar ring
of SPRC-33 is not noticeable in the SDSS
images, however, we included the galaxy in the list, since its
outer ring reveals both HI and a young stellar population
(see Section \ref{sect_known}). At the same time, our catalogue
does not include the NGC~6822 and SDSS J102819.24+623502.6
galaxies, mentioned in the introduction, since their HI polar discs
revealed no detected stellar population.

\subsection{The best candidates}

We classified  70 galaxies  into this type, the belonging of  which
to the PRG class is almost undoubted, and their images are similar
to those, observed in the ``classical'' long-known PRGs, like
NGC4650A, NGC~2685 or UGC~7576. In the galaxies of this group,
the outer component is extended,
homogeneous, often distinguished by a bluer colour in comparison
to the central galaxy, indicating the presence of a young stellar
population. Sometimes the rings are accompanied by a dust belt,
crossing the central galaxy along the minor axis: SPRC-1, SPRC-25,
SPRC-27, SPRC-42, SPRC-48, SPRC-54, SPRC-66, SPRC-69. In most
candidates the outer ring is close to being viewed edge-on,
which is obviously due to the selection effect during the visual
selection of objects. In some cases, we can clearly see how the
outer ring, inclined to the line of sight covers the central body:
SPRC-39, SPRC-47, SPRC-51, SPRC-56, SPRC-65, SPRC-65,  SPRC-69.

Detailed photometric studies of the SPRC-1 and SPRC-41 galaxies,
which argue for their membership in the PRG class are presented in
the papers by  \citet{Resh2011} and  \citet{Finkelman2011},
respectively. In the latter case, the authors used the SUBARU Deep
Field images, where the outer component looks considerably more
extended than in the SDSS images. The results of new spectral
observations for 5 of our best candidates are presented in
Sections \ref{sect_known} and \ref{sect_obs}.

\begin{figure*}
\includegraphics[width=\textwidth]{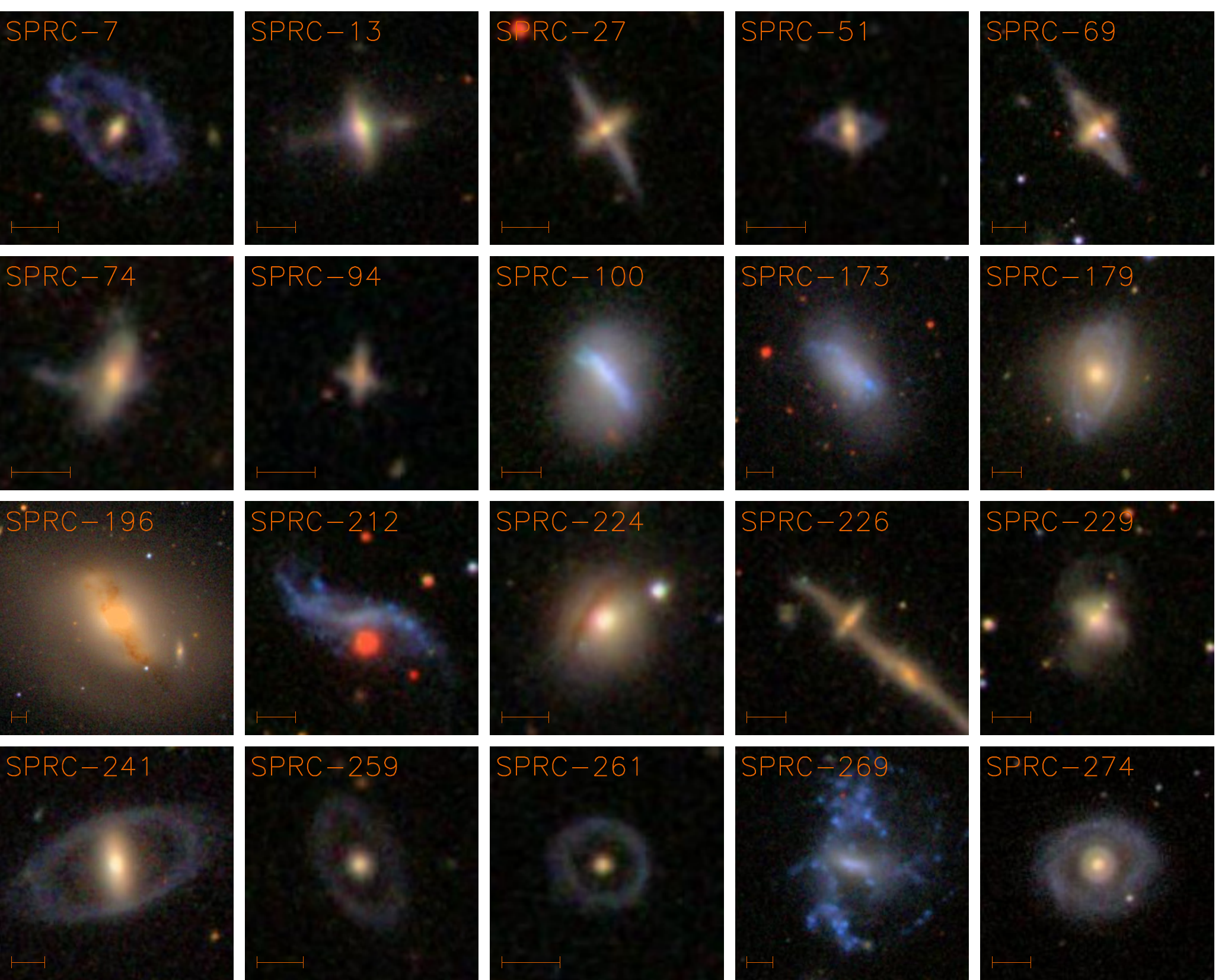}
\caption{ Examples of PRG candidates --- a color combination of
images in the $g,r,i$ provided by  SDSS server. Different
types of objects are represented (from top to bottom: the best
candidates, good candidates, related objects, and possible
face-on rings). The scale bar corresponds to the angular size of
10 arcsec.} \label{fig_example}
\end{figure*}

\begin{figure*}
\includegraphics[width=\textwidth]{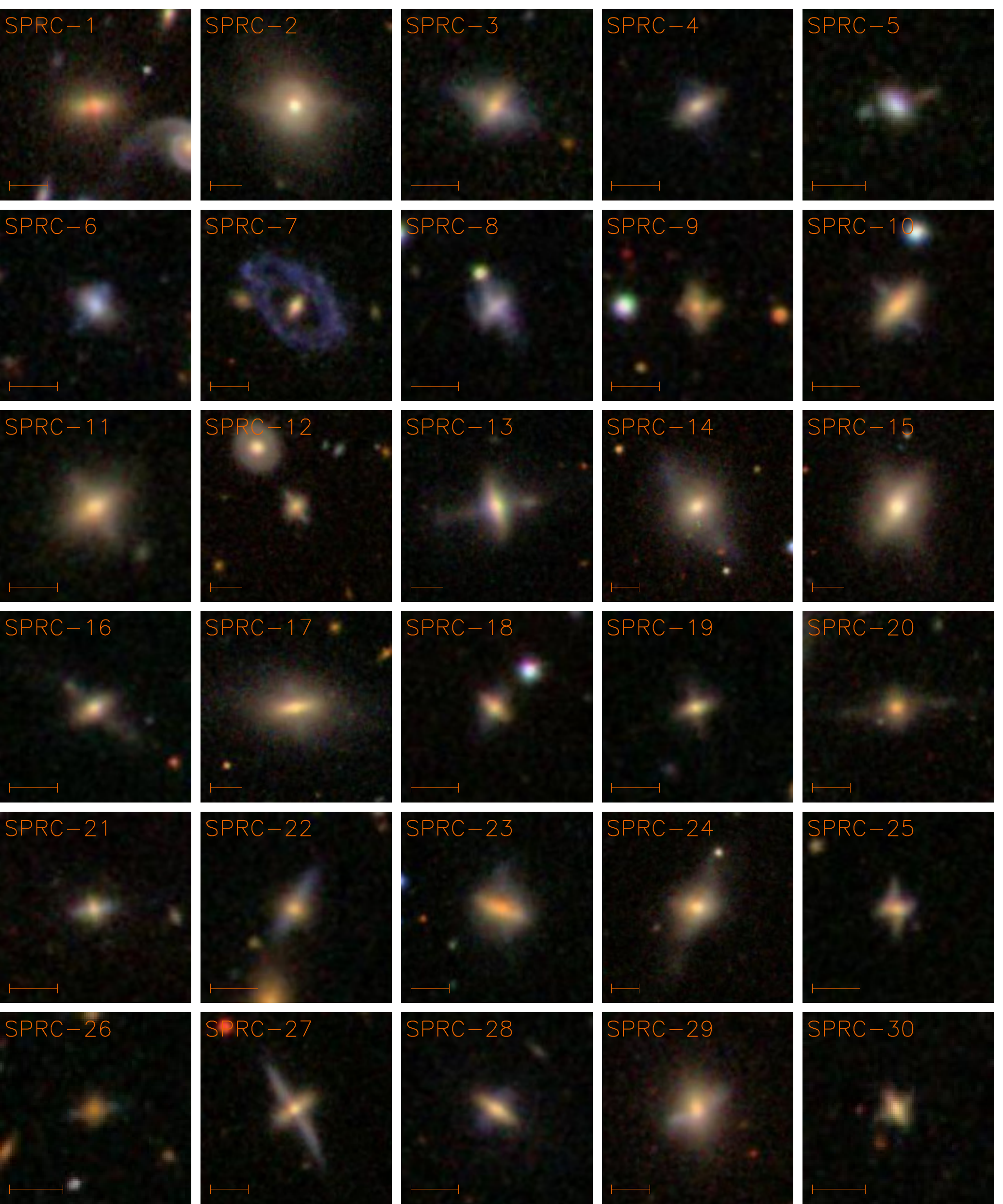}
\caption{Optical images of all the galaxies of our catalog, a
combination of images in the $g,r,i$ filters according to the SDSS
DR8 data. The  scale bar corresponds to the angular size of
$10$ arcsec.} \label{fig_atlas}
\end{figure*}

\begin{figure*}
\includegraphics[width=\textwidth]{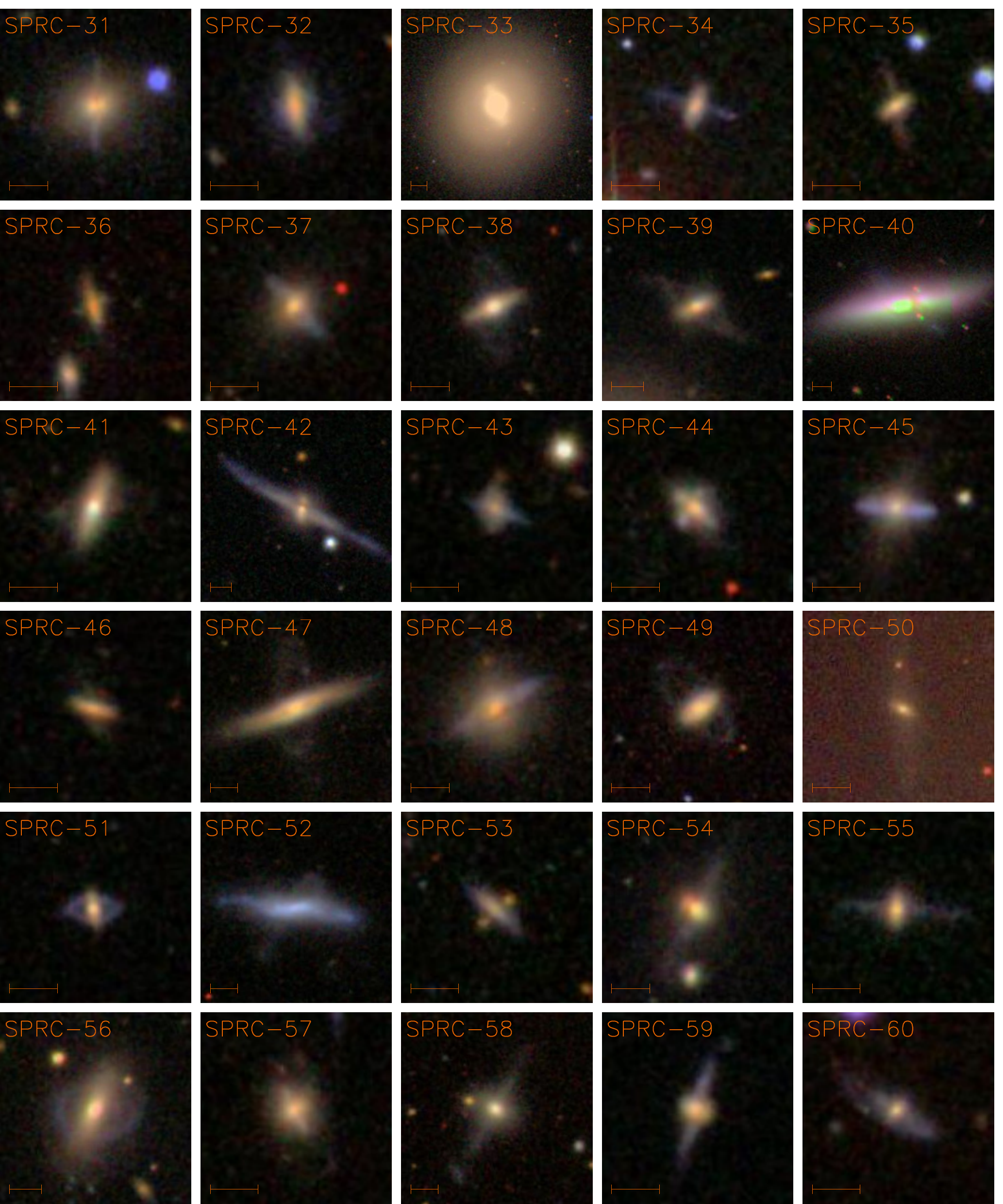}
\contcaption{}
\end{figure*}

\begin{figure*}
\includegraphics[width=\textwidth]{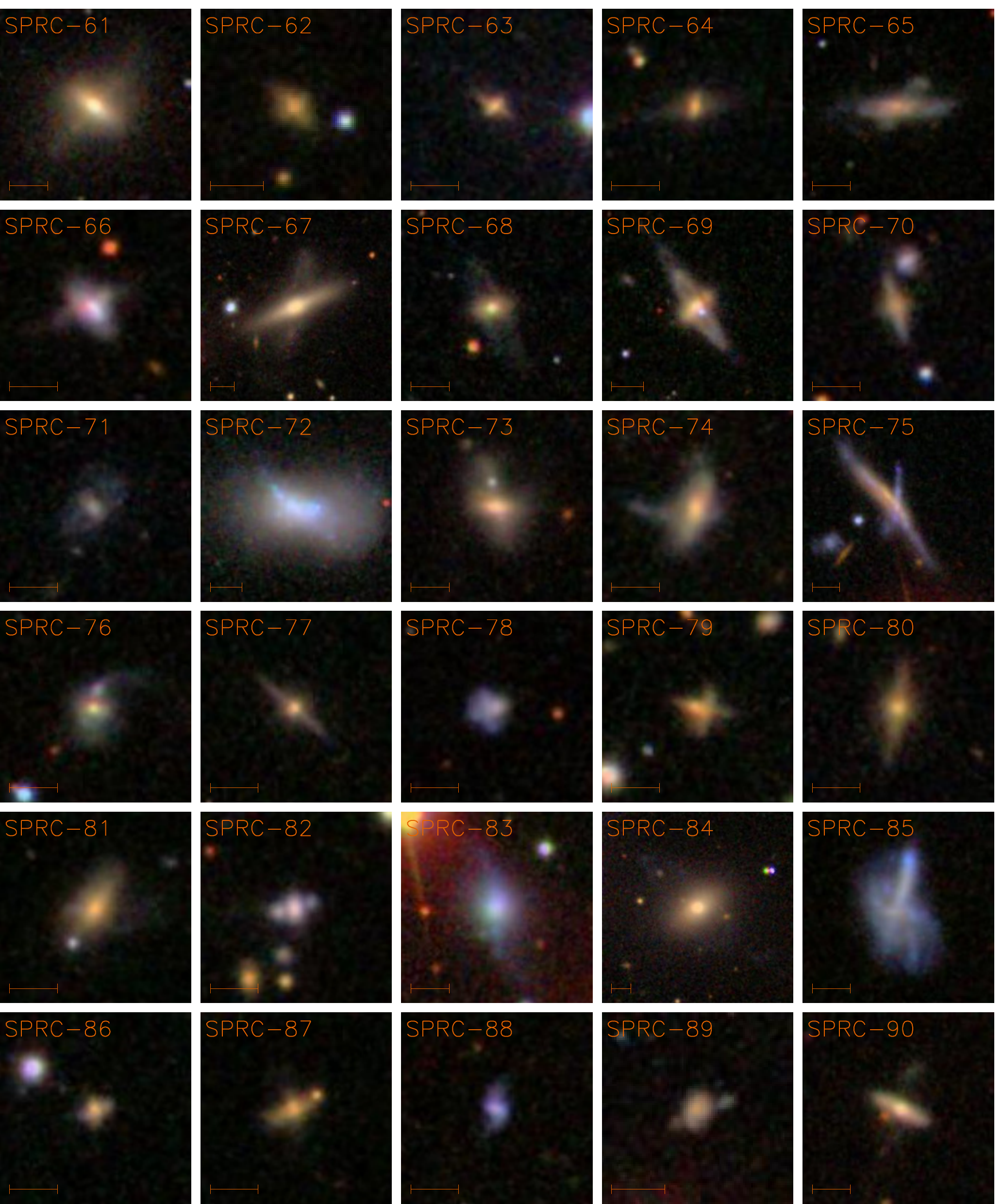}
\contcaption{}
\end{figure*}

\begin{figure*}
\includegraphics[width=\textwidth]{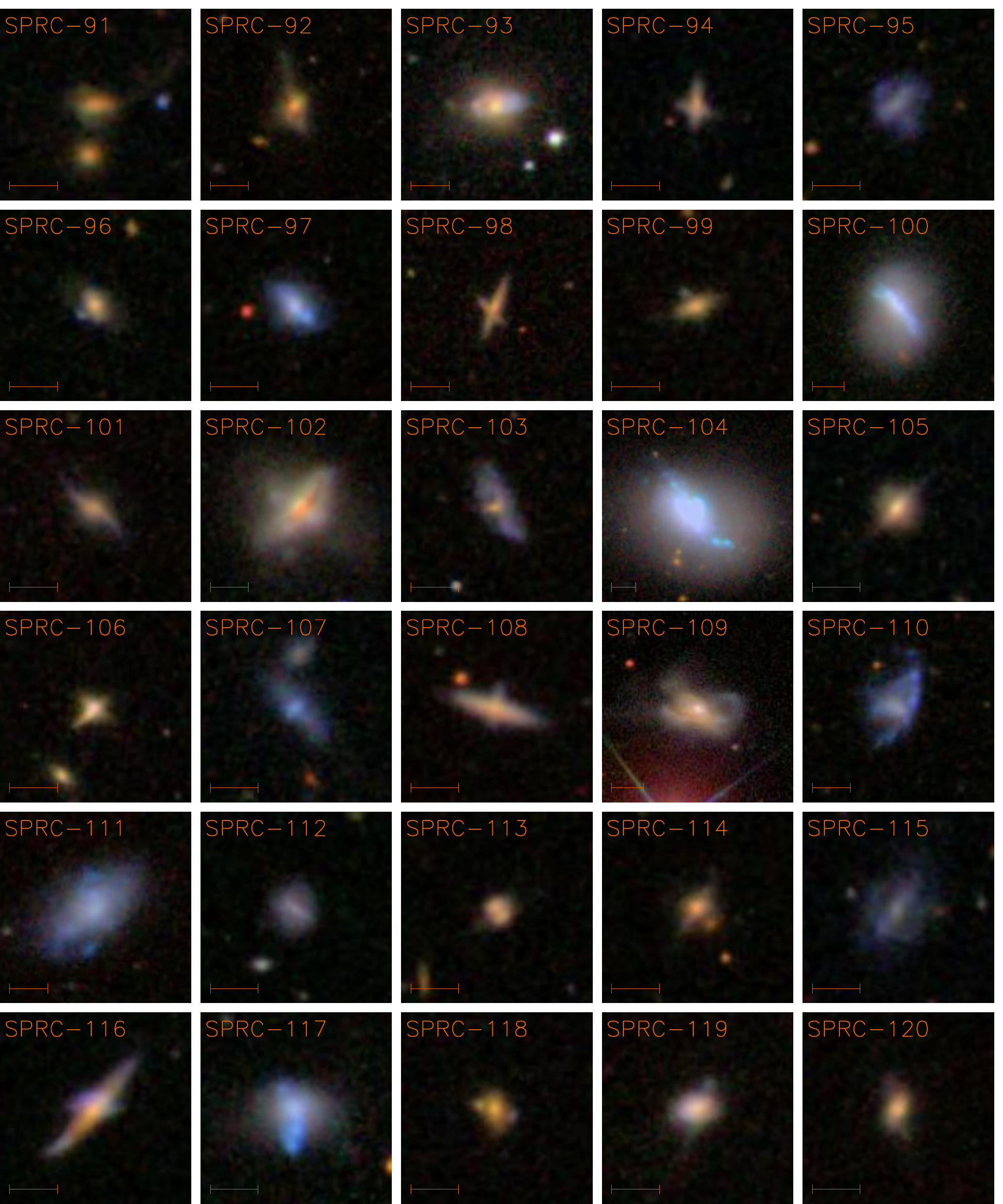}
\contcaption{}
\end{figure*}

\begin{figure*}
\includegraphics[width=\textwidth]{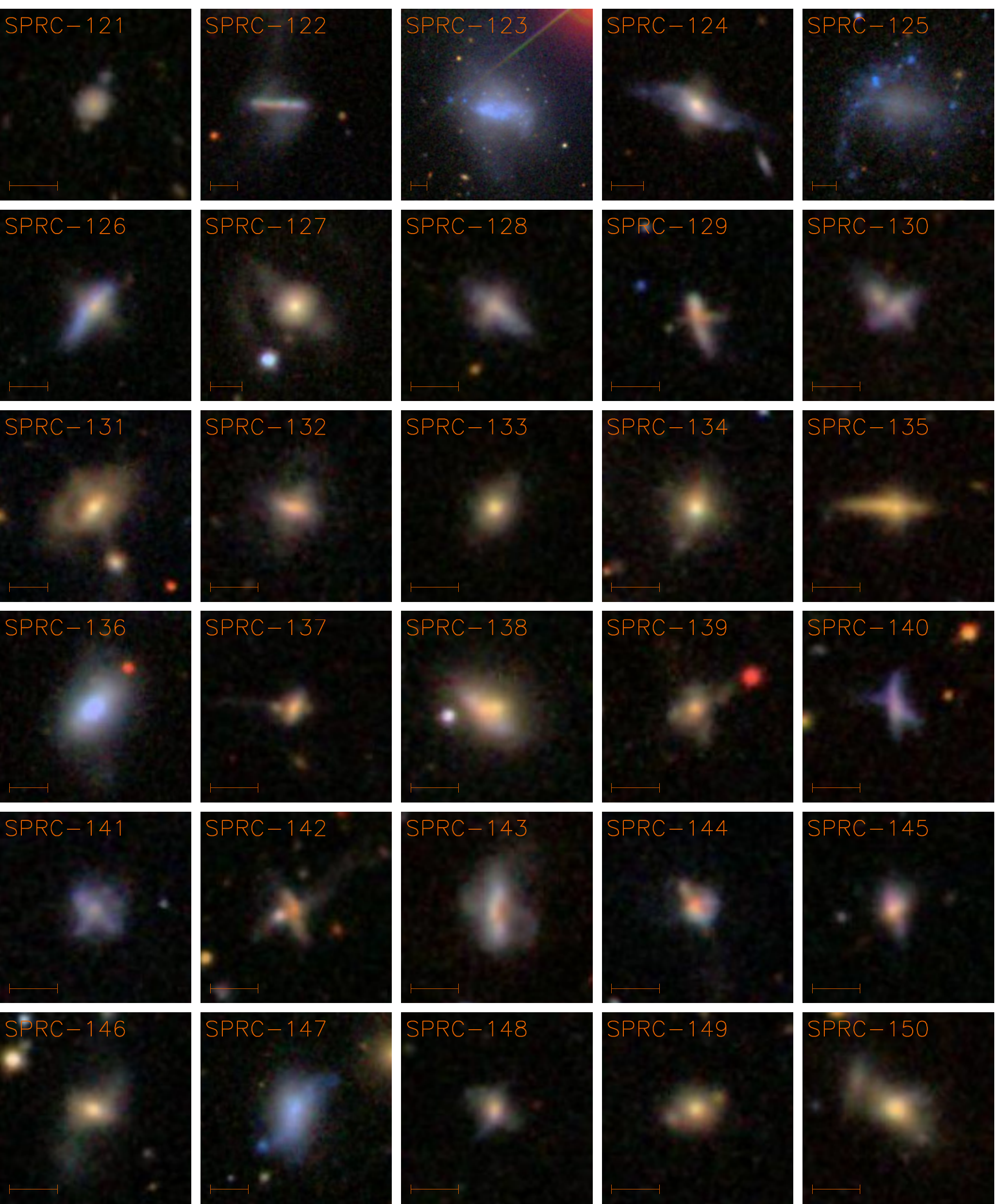}
\contcaption{}
\end{figure*}

\begin{figure*}
\includegraphics[width=\textwidth]{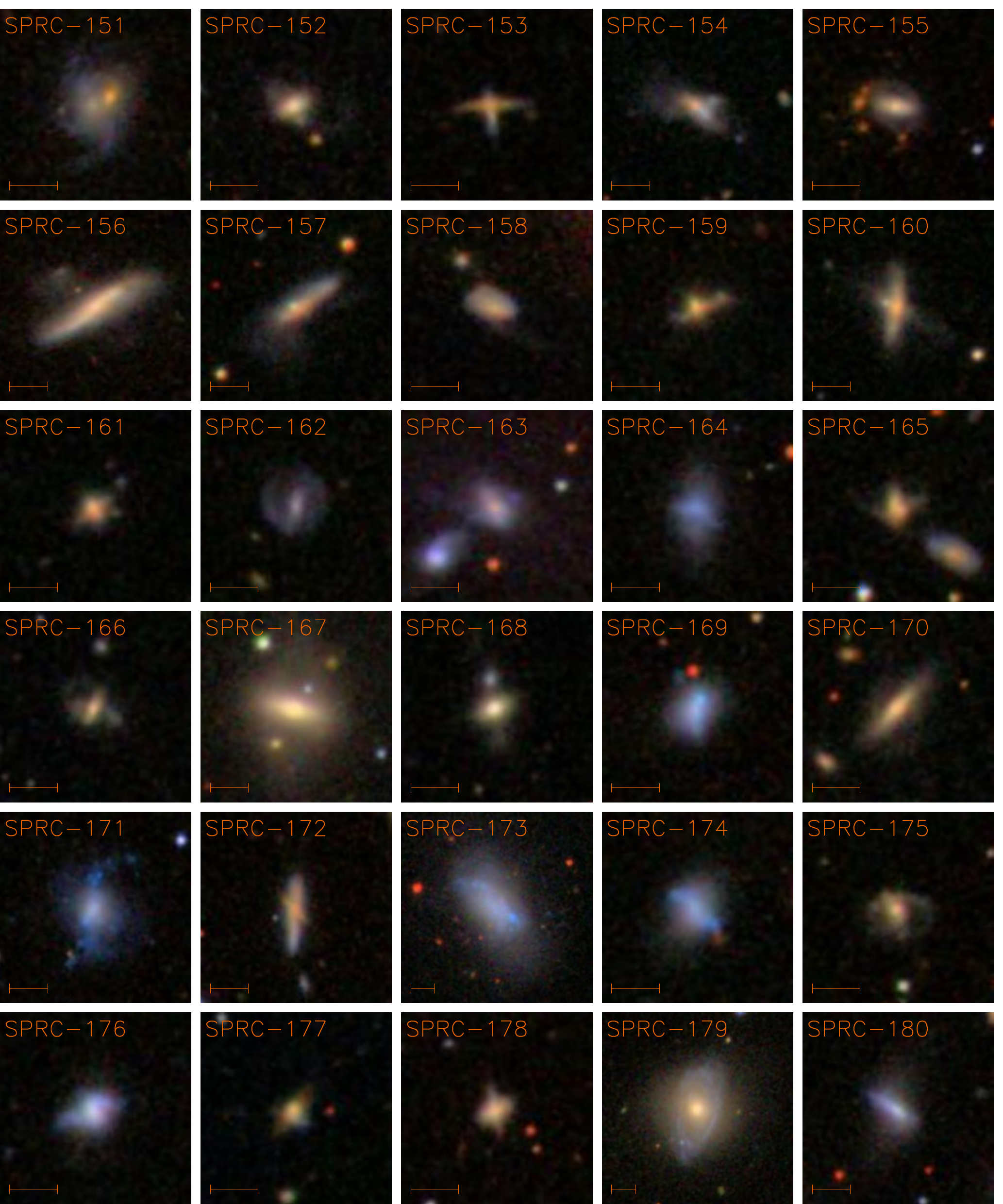}
\contcaption{}
\end{figure*}

\begin{figure*}
\includegraphics[width=\textwidth]{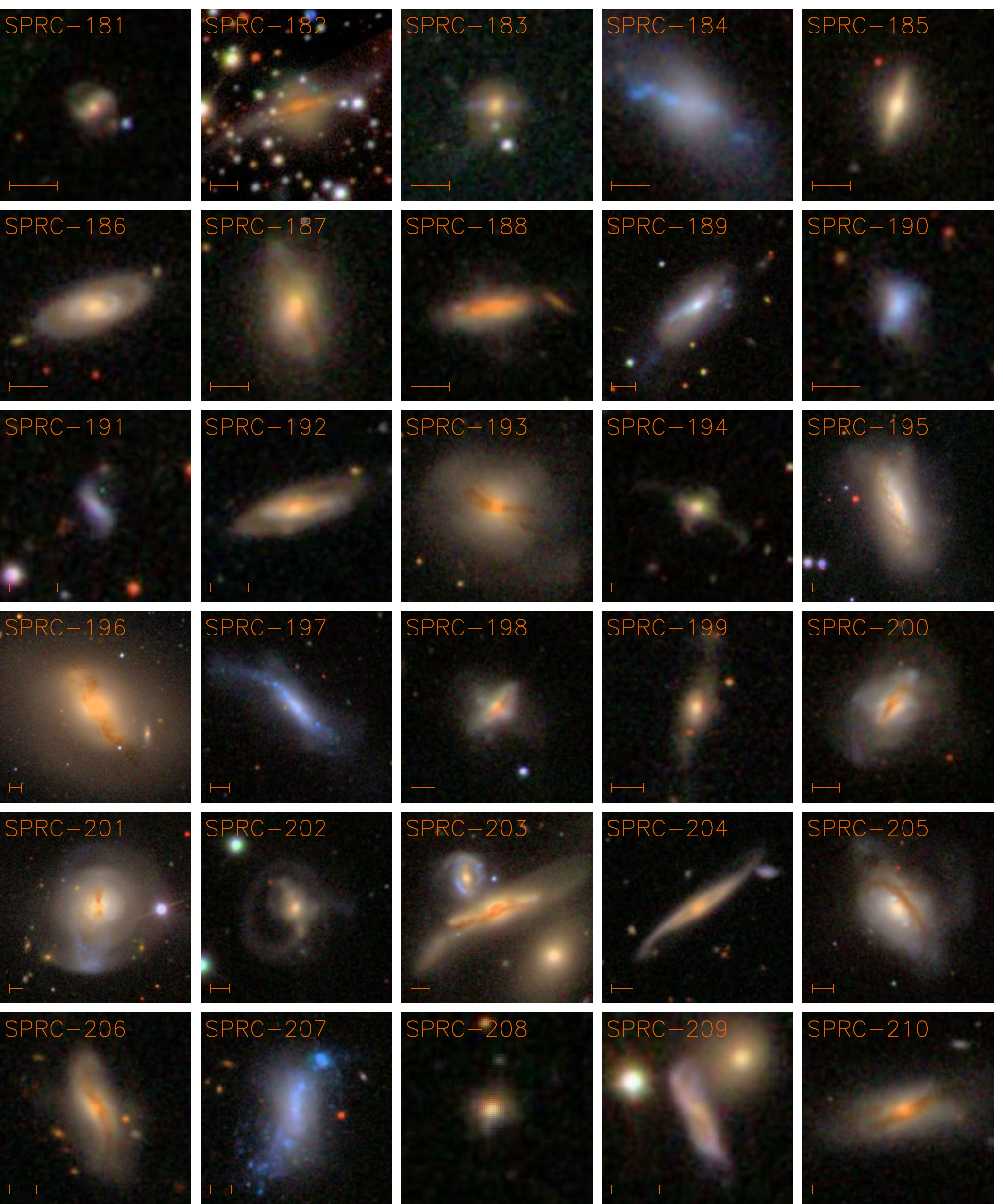}
\contcaption{}
\end{figure*}

\begin{figure*}
\includegraphics[width=\textwidth]{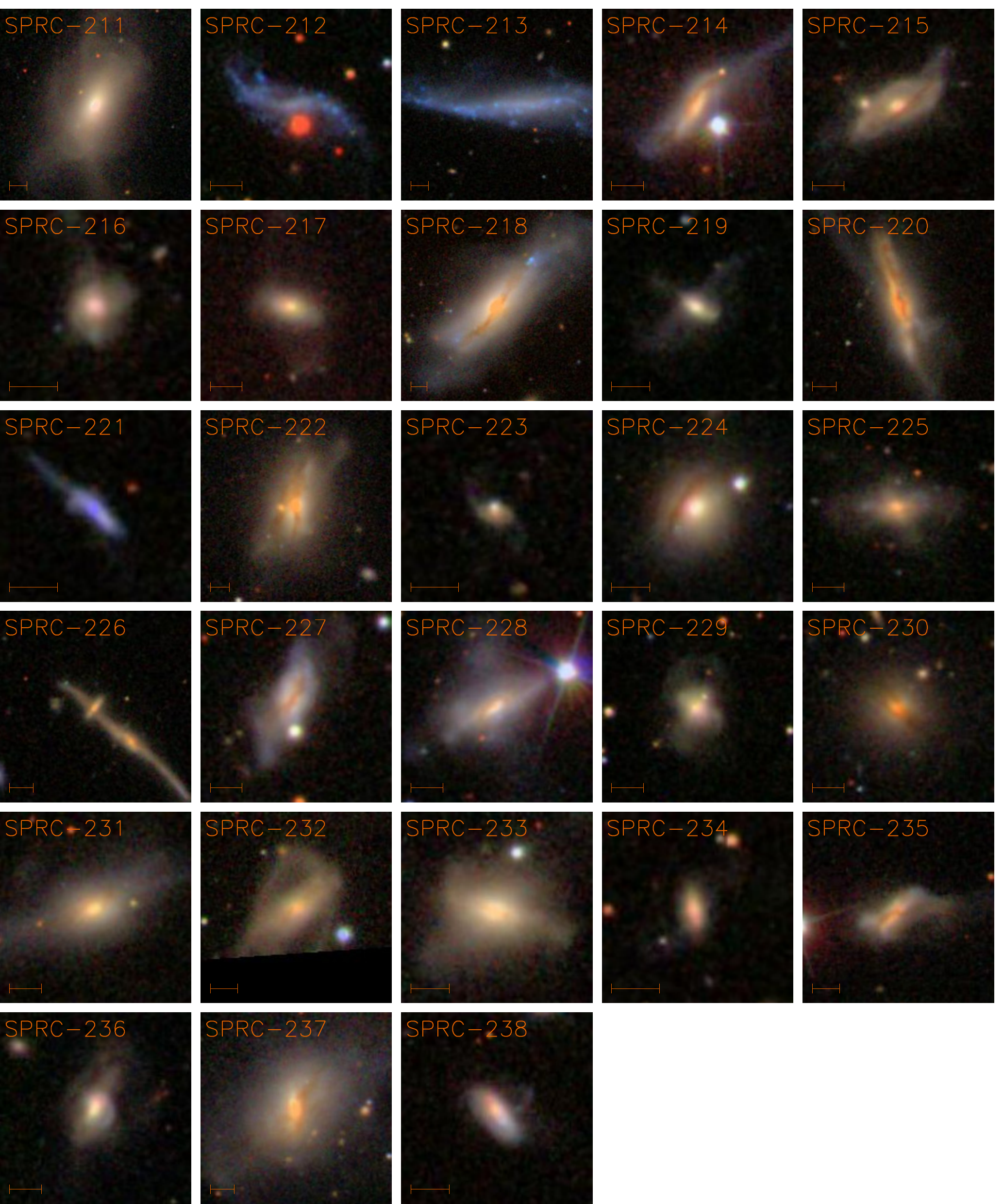}
\contcaption{}
\end{figure*}

\begin{figure*}
\includegraphics[width=\textwidth]{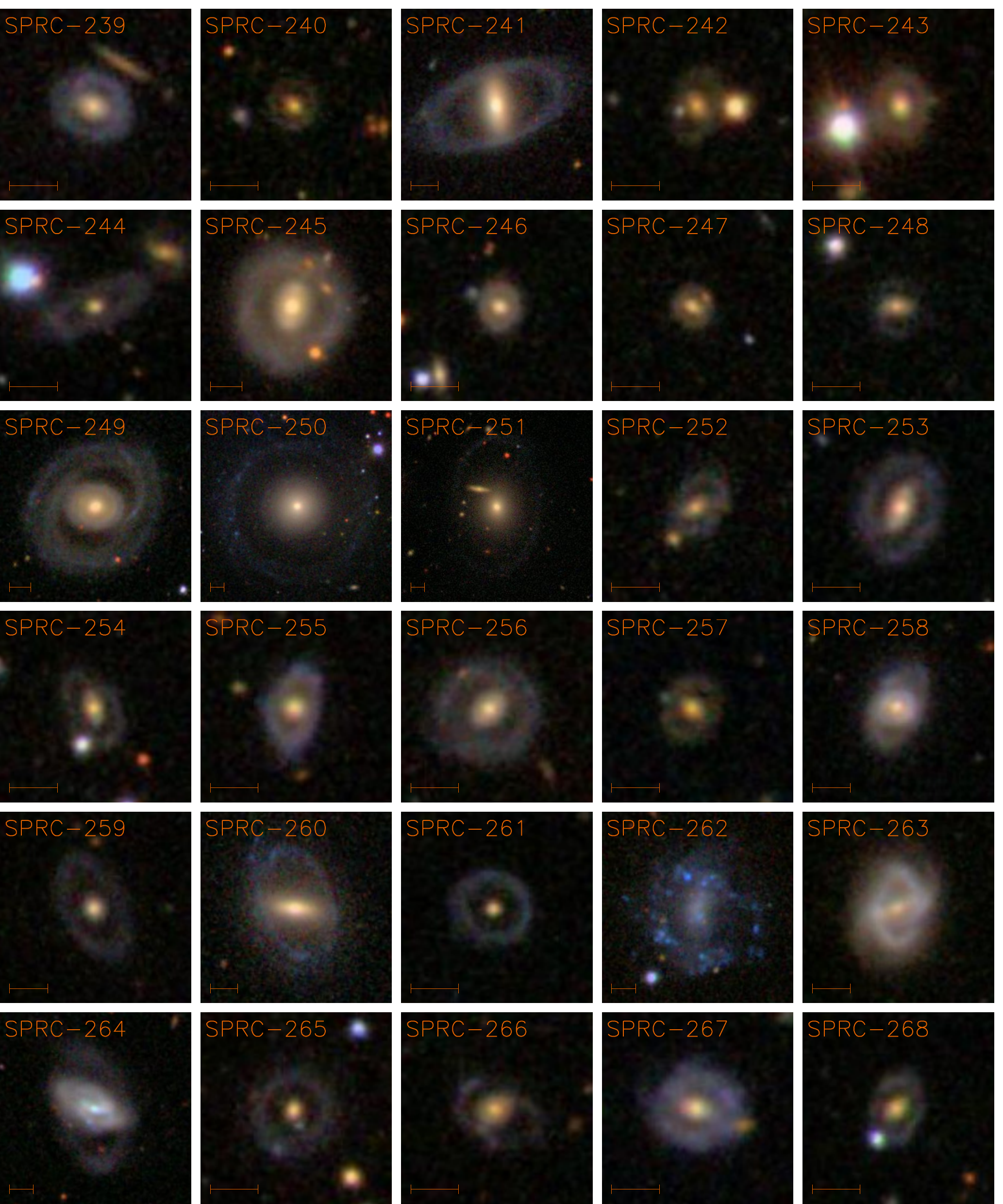}
\contcaption{}
\end{figure*}

\begin{figure*}
\includegraphics[width=\textwidth]{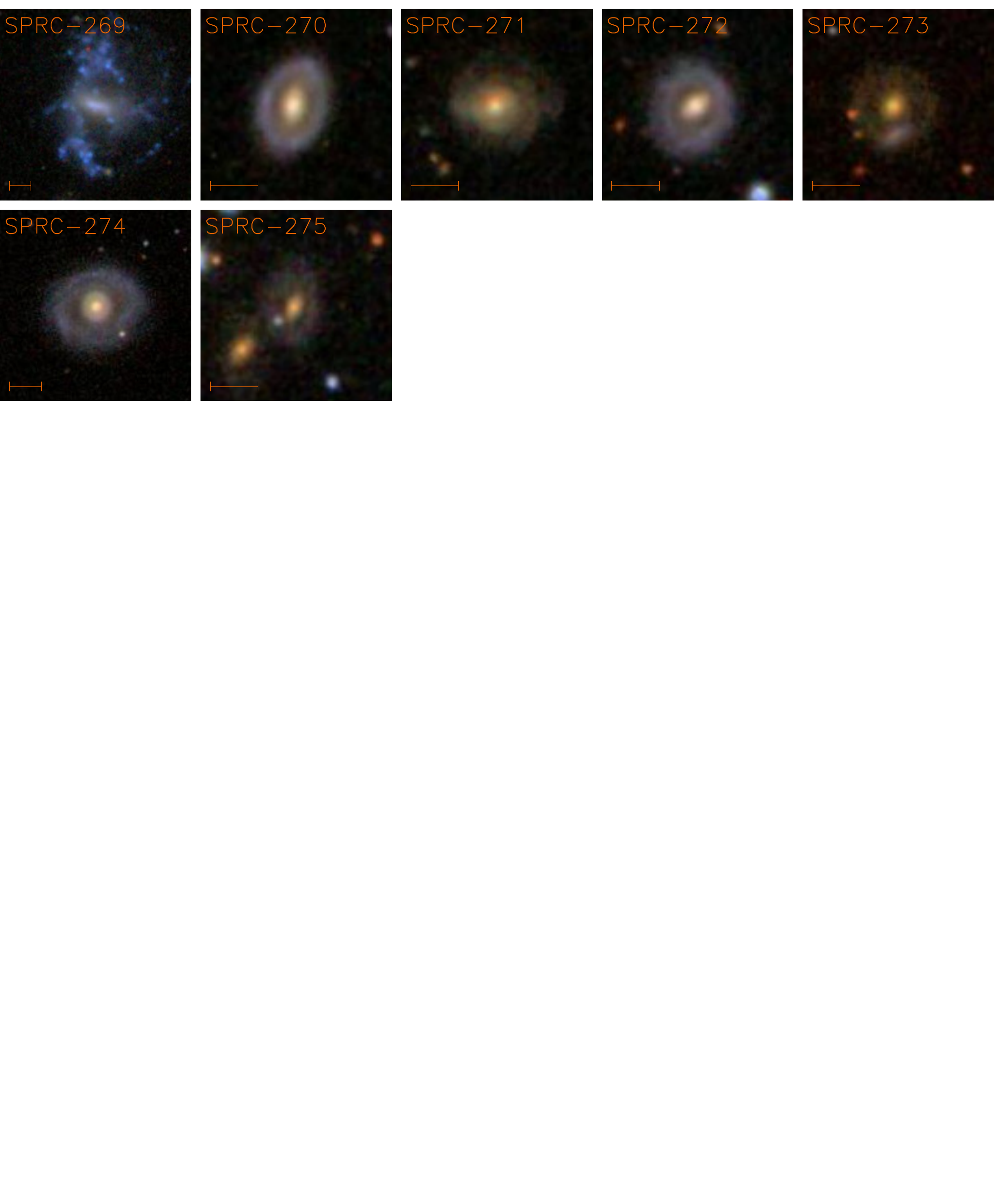}
\contcaption{}
\end{figure*}

\subsection{Good candidates}

In this largest group  (115 objects)  we included the candidates
which are likely to have a polar component, but their shapes are
slightly different from  the classical PRGs.   Main features of
the galaxies in this group are:

\begin{itemize}
\item In the majority of candidates the outer component is
inhomogeneous in terms of brightness, sometimes it looks like an
unclosed arc, or reveals separate fragments of the ring.

\item  The blue ring is lost in the background of the central
galaxy,   having a larger  size: SPRC-100, SPRC-104, SPRC-111, SPRC-117,
SPRC-123, SPRC-169, SPRC-173, SPRC-174, SPRC-179. It is  possible
that in  some of these galaxies we are dealing with a random distribution of the HII regions.

\item Both components are seen nearly edge-on, but do not always
intersect exactly in the centre: possible random
projections of two disc galaxies. Unfortunately, small angular
sizes of objects do not allow making any firm conclusions.
Accurate measurements of the systemic  velocity of each
component are required here. The most typical examples are:
SPRC-75, SPRC-79, SPRC-82, SPRC-89, SPRC-94, SPRC-98, SPRC-106, SPRC-116, SPRC-124, SPRC-129, SPRC-130, SPRC-140, SPRC-142, SPRC-153, SPRC-160.
\end{itemize}

We intentionally left  one misclassified  galaxy in our catalogue as a good
illustration of the difficulties faced in the process of visual
selection of candidates. Based on morphology, the SPRC-178 galaxy
seemed to be a good PRG candidate, but spectral observations have
shown that it is simply a pair of interacting galaxies (see
Section \ref{sect_obs}).

\subsection{Related objects}

This group consists of  53 galaxies, the images of which suggest
that although a part of the matter in the outer regions  is not
accumulated in exactly one plane, it is most likely rotating
outside the plane of the main disc.  Typically, these are the
systems that undergo different dynamical states of  interaction
before they relax. Galaxies with  the following description fall into
this group:

\begin{itemize}
\item Galaxies with strong warps of their stellar disks, typical examples:
SPRC-191, SPRC-192, SPRC-197, SPRC-203, SPRC-212, SPRC-214,
SPRC-215, SPRC-218. It is possible that some of them are similar
to the well-known PRG NGC~660 \citep{vanDriel1995}.

\item Galaxies with the stellar discs viewed almost face-on,
their centres are crossed by a dust belt, which is often curved:
SPRC-193, SPRC-195,  SPRC-196, SPRC-200, SPRC-205, SPRC-222,
SPRC-224, SPRC-230, SPRC-237. The closest analogue among the PRC
catalogue  objects is NGC~5128 (Cen A).

\item Interacting systems in which the matter of a destroyed
companion is presumably distributed far off the main plane of the galaxy. The
most typical examples: SPRC-194, SPRC-201, SPRC-211, SPRC-219,
SPRC-228, SPRC-236.
\end{itemize}

The galaxy SPRC-196 (NGC~2968) has already been discussed
previously by \citet{Sarzi00}, who noted that this galaxy is
related to PRGs, as the bulge isophotes are elongated
perpendicular to the disc. For SPRC-201, the literature has
evidence of the existence of a kinematically-isolated subsystem
(see Section \ref{sect_known}).

\subsection{Possible face-on rings}

It was  previously noted that among PRGs there should exist plenty of
objects with their polar structures moderately inclined  to the line-of-sight (nearly face-on),
disguising themselves as ``normal'' ring galaxies
with bars \citep{Whitmore1990, Resh2011}.  To date, we already
know of two galaxies with polar rings,  slightly   inclined to the plane of the sky: ESO 235-58 \citep{ButaCrocer1993}, and the object
from our catalogue SPRC-7 \citep{Brosch2010}. The last category
includes  37 galaxies, the images of which are slightly different
from what is normally expected from the outer rings of galaxies
with central bars  \citep[see, e.g.,][]{ButaCombes1996}. As a rule,
here we do not observe any disc components between the bar and the ring, and sometimes the bar has
a surprisingly low ellipticity, like in  SPRC-239, SPRC-244, SPRC-255,
SPRC-259, SPRC-261,  SPRC-261,  SPRC-265, SPRC-267, SPRC-274.   Some of these galaxies seem to be very
similar to well-known Hoag's object (PRC D-41) which consists of an elliptical galaxy surrounded by blue ring of young stars and ionized gas.
 \citet{Whitmore1990} considered this galaxy as possibly related with PRGs.  The most similar galaxies in our catalogue are SPRC-250 \citep[see its photometric study in ][]{FinkelmanBrosch2011},
SPRC-261 and SPRC-265.

Of special note is the object SPRC-269, the companion of which is
also included in our catalogue -- SPRC-136. We might well see here
the first system of two PRGs!

\begin{figure*}
\centerline{
\includegraphics[width=0.2\textwidth]{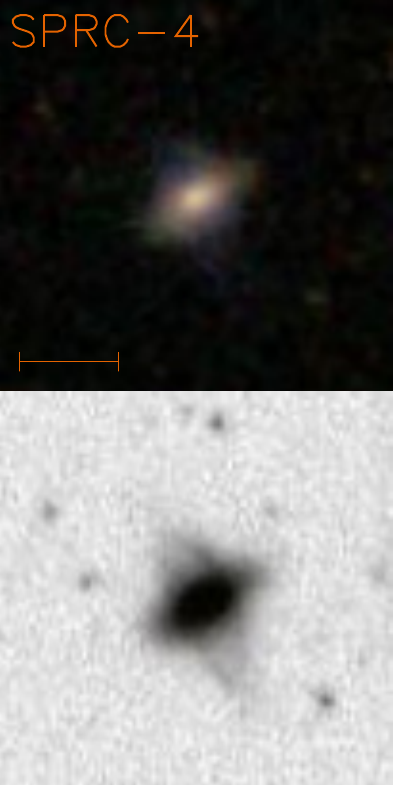}
\includegraphics[width=0.2\textwidth]{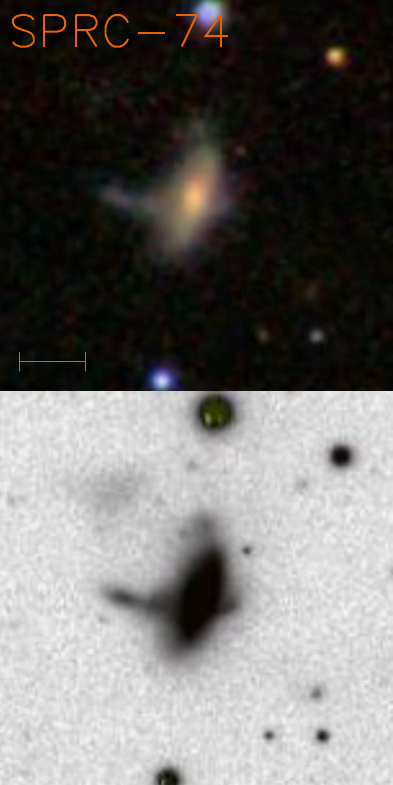}
\includegraphics[width=0.2\textwidth]{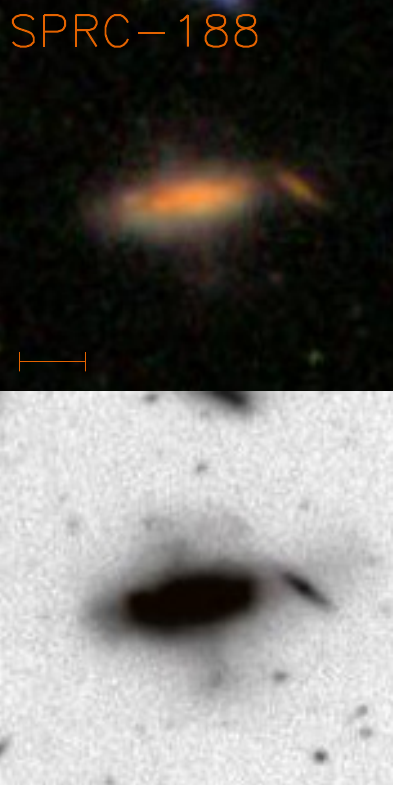}
\includegraphics[width=0.2\textwidth]{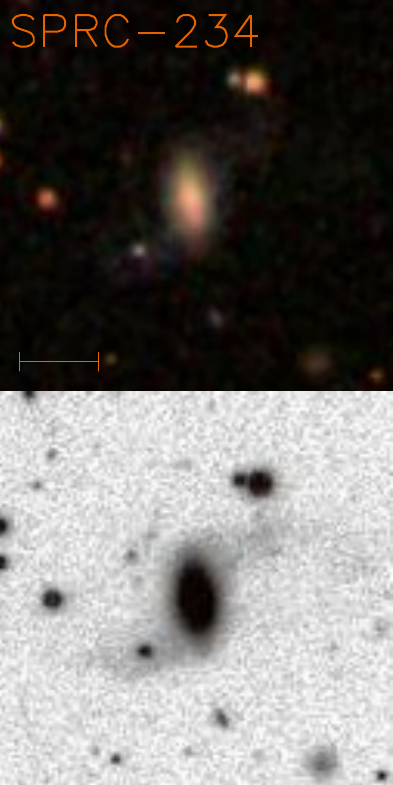}
\includegraphics[width=0.2\textwidth]{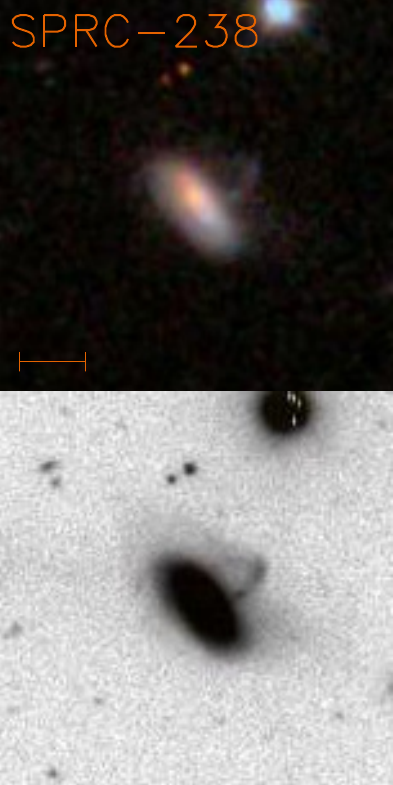}
}
\caption{ Examples of deep Stripe82 $r$-band monochromatic images of the SPRC galaxies (bottom  panels) together with their standard multicolour
SDSS/DR8 images (top panels). The scale bar corresponds to the angular size of
10 arcsec.}
\label{fig_stripe82}
\end{figure*}

\subsection{Stripe82 deep images}
\label{sect_stripe82}

There are  13 SPRC galaxies in the  SDSS/DR7 ``Stripe82'', a 120$^\circ$ long, 2.5$^\circ$ wide stripe along
the celestial equator that has been imaged multiple times. The Stripe82 imaging reaches $\sim2$ mag deeper
than single SDSS scans \citep{sdss7}.
Fig.~\ref{fig_stripe82} provides a good illustration how the Stripe82 data enhance the detection limit for faint extended features around the galaxies from our catalogue. For example, the external blue ring has a very low contrast on the standard SDSS image of   SPRC-4, although the Stripe82 deep image reveals  this feature in detail (Fig.~\ref{fig_stripe82}).

Obviously,  a  fraction  of low surface brightness polar rings which are hidden  in standard SDSS images could be detected via  the Stripe82 data. Inspecting  galactic images in the Stripe82 field we can estimate a possible fraction of such low-brightness structures, similarly to what was done by \citet{Kaviraj2010}, who studied the signatures of recent interactions in a sample of early-type galaxies. The detailed analysis of the low-contrast PRG candidates is beyond the scope of the current paper, however, it may be interesting in the future.

\begin{table*}
\caption{Log of the observations}\label{tab_obs}
\begin{tabular}{lrrccc}
\hline
Galaxy    & Slit $PA$ & Date             & Sp. range        &  Exp. time             &  Seeing  \\
(SPRC )  &   (deg)&                       &  (\AA)               &     (s)                       &   (arcsec)\\
\hline
SPRC-10        &  39       & 07.12.2010  &    3700--7290 &    $4\times1200$ &   1.6\\
               &  138      & 07.12.2010  &    3700--7290 &    $5\times1200$ &   1.5\\
SPRC-14        &  35       & 08.12.2010  &    3700--7290 &    $4\times900$ &   2.2\\
               &  130      & 08.12.2010  &    3700--7290 &    $4\times900$ &   2.1\\
SPRC-39        &  53       & 08.12.2010  &    3700--7290 &    $5\times900$  &   2.3\\
SPRC-60        &  56       & 11.09.2010  &    4800--8480 &    $4\times1200$ &   1.0\\
SPRC-69        &  34       & 31.08.2010  &    4050--5830 &    $6\times1200$ &   1.4\\
               &  130      & 31.08.2010  &    4050--5830 &    $4\times1200$ &   2.1\\
SPRC-178       &  12       & 10.09.2010  &    4800--8480 &    $5\times1200$ &   1.6\\
               &  114      & 11.09.2010  &    4800--8480 &    $3\times1200$ &   1.5\\
  \hline
\end{tabular}
\end{table*}

\section{Kinematical confirmation of PRG candidates}
\label{sect_obs}

 In this section we summarize the available in the literature kinematical data for the SPRC objects, as far as
we present the results of original observations of six previously uninvestigated galaxies.
The results of observations clearly demonstrate  that morphological selection is an effective
tool for recognizing  the true PRGs --- 5 of 6 objects were confirmed to be classical PRGs.

\subsection{Kinematical data from previous studies}
\label{sect_known}

\textbf{SPRC-7}.
The kinematics of its gaseous and stellar subsystems was studied by
\citet{Brosch2010}. It is demonstrated that the central object is
an early-type galaxy. The outer ring, characterised by its
blue colour, consists of a younger stellar population and emits lines of
ionized gas. The analysis of the velocity field in the
H$\beta$ line  has shown that this giant ring (with a diameter of
48 kpc) rotates at a noticeable angle to the plane of the central
galaxy. According to various estimates, the angle between them
amounts to $\Delta i=58\pm10\degr$ or $73\pm11\degr$.

\textbf{SPRC-33 (NGC~4262)}. In a recent paper by
\citet{Bettoni2010} it was shown that both the ionized gas in the
inner region and the neutral hydrogen at large distances from the
centre, rotate in the plane,  that is strongly inclined to the stellar disc
of this galaxy with a central bar. In
addition, the UV images demonstrate the presence of a young
stellar population in the ring of neutral hydrogen, having the
outer diameter of about $18-20$ kpc.  \citet{Bettoni2010}
chose to speak of an inclined ring, but a formal calculation of
the mutual inclination angle \citep[see formula (1) in][]{Moiseev2008}
yields two solutions: $\Delta  i=39\degr$ and
$90\degr$. The latter corresponds to the polar ring, as noted in
their subsequent work \citep{Buson2011}.

\textbf{SPRC-40 (NGC~5014).} On the background of a bright
lenticular body of the galaxy a small blue ring is visible with a
diameter of approximately 45 arcsec ($\sim3.9$ kpc), strongly
inclined to the plane of the galaxy, elongated along
$PA\approx45\degr$. According to the radio observations of
\citet{Noordermeer2005} with the WSRT, the HI rotation occurs in
the disc, the kinematic axis of which coincides with this blue ring,
and its diameter reaches about 15 kpc. The HI distribution also
reveals a tidal tail, extending over at least 20 kpc. It is noted
that the galaxy belongs to a group, rich in HI clouds, the
accretion of which (or a capture of a companion) has formed a
polar structure.

\textbf{SPRC-67 (CGCG 225-097, PGC~060020)}. The  kinematics
and photometry of this galaxy were studied at the 6-m telescope
by \citet{Karataeva2011}. It was found that the outer ring  rotates at
a large angle to the plane of the disc of the central
galaxy, therefore the polar ring is kinematically confirmed.

\textbf{SPRC-201 (NGC~3656).} A well-known interacting galaxy.
According to \citet{Young2002} the molecular gas here is rotating
in a warped inclined disc with a diameter of at least $8-9$ kpc.
Furthermore, \citet{Balcells1990} have discovered  at the distance
of $r<1$ kpc a kinematically decoupled core, rotating perpendicular
to the disc of the galaxy.

\subsection{6-m telescope observations}

Spectral observations were carried out at the prime focus of the SAO RAS
6-m telescope. In August 2010 we used the SCORPIO focal
reducer \citep{AfanasievMoiseev2005} with a EEV CCD42-40 CCD chip
sized $2048\times2048$ pixels as a detector. The further
observations were carried out with a new experimental instrument
SCORPIO-2 \citep{AfanasievMoiseev2011}, where a E2V CCD42-90 CCD chip, sized
$2048\times4600$ was implemented, providing a two times larger
spectral range. The other characteristics of both devices were
similar -- the slit width of 1 arcsec, its height -- 6 arcmin,
the scale along the slit amounted to 0.35 arcsec/pix, and the
spectral resolution $FWHM=5$\AA. The log of observations is
presented in Table~\ref{tab_obs}. The spectral range included both
the   absorption lines of the stellar population, and emission lines of the ionized gas: H$\beta$
and [OIII] for the August set  of observations, \Ha\, and [NII] -- in other cases.

The processing of spectra  was done using the codes we wrote in the
IDL environment \cite[see, e.g.,][]{Zasov2008}. The measurements
of the distribution of line-of-sight velocities and stellar velocity
dispersion  were done by cross-correlating the spectra of galaxies
with the spectra of the template star
observed in the same night. The measurement technique is described
in our previous papers \citep{Moiseev2001,Zasov2008}. We observed
several template stars belonging to the spectral types III~G8 --
III~K3. For the final measurements we were selecting a template,
giving a maximum correlation coefficient. Line-of-sight velocities of
ionized gas were measured via the Gaussian approximations of the
selected emission line profiles.

In this paper we only briefly discuss the results obtained, mainly
confining ourselves to the confirmation/disproof of the presence
of a kinematically isolated polar component. In the subsequent
papers we would like to further explore the characteristics of the
gas and stars in these galaxies, the composition of their stellar
population, rotation curves, etc.

\subsection{Kinematics of gas and stars}

For each candidate, we obtained the spectrum in two position
angles, oriented along the major axis of the inner body and the
possible polar ring. SPRC-38 and SPRC-58 are the exceptions,
observed only in one   slit position.   Fig.~\ref{fig_spectr1}
 demonstrates  the distribution of line-of-sight
velocities of the stars and ionized gas (from the brightest
emission lines) along each  slit position.  The kinematic components,
corresponding to the central galaxy and the outer structure are
also marked  in the left panels on the figure. From now on, we will refer to them as
``the central disc'' and ``the polar ring''.

\begin{figure*}
\includegraphics[width=0.97\textwidth]{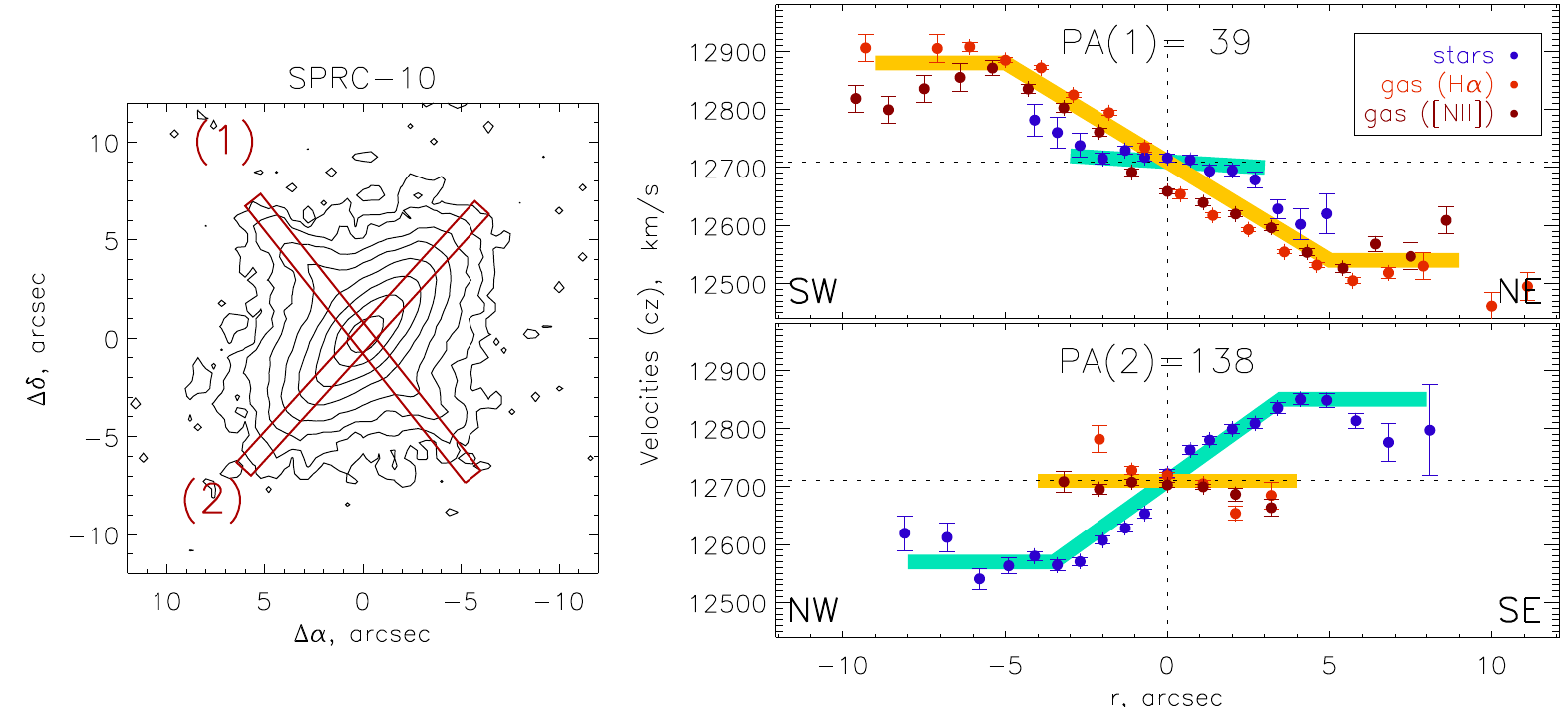}
\includegraphics[width=0.97\textwidth]{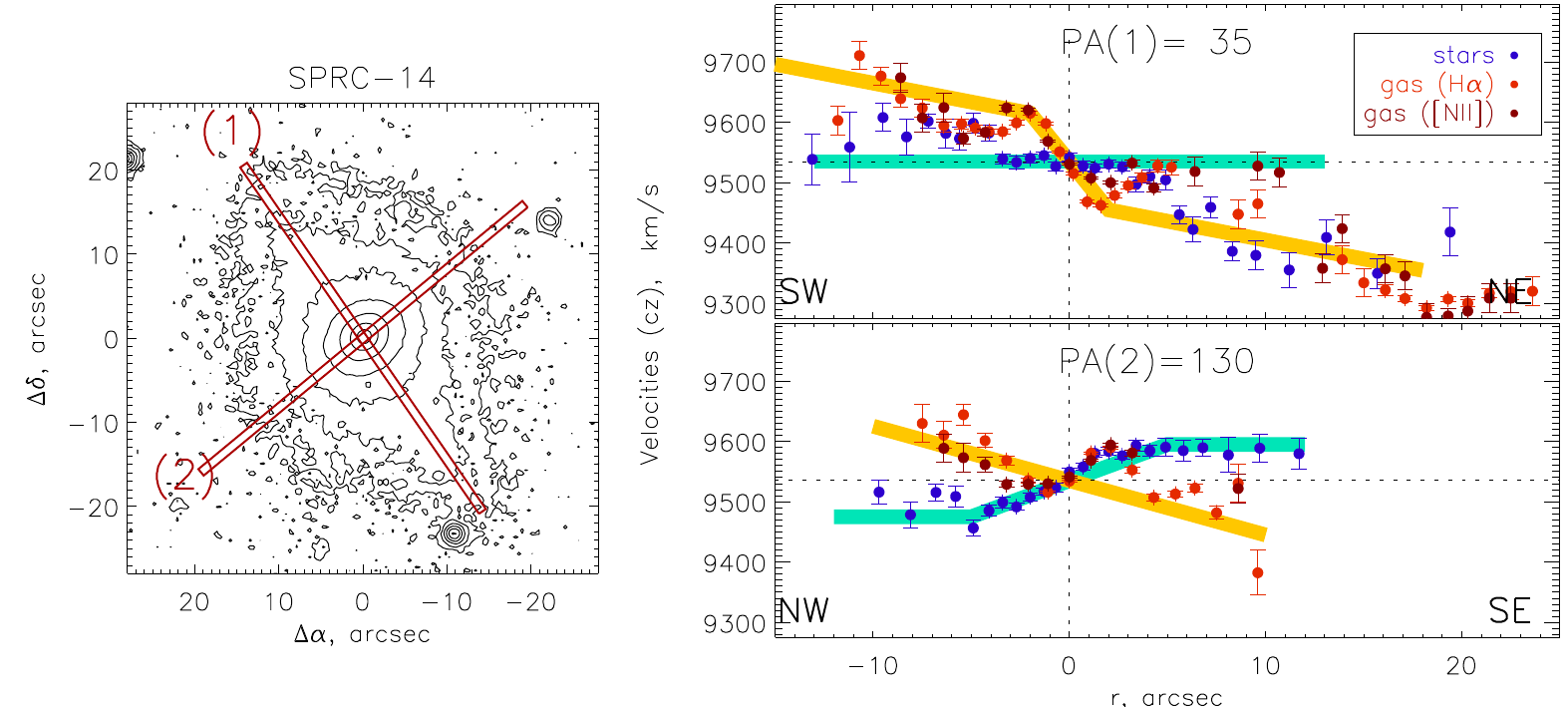}
\includegraphics[width=0.97\textwidth]{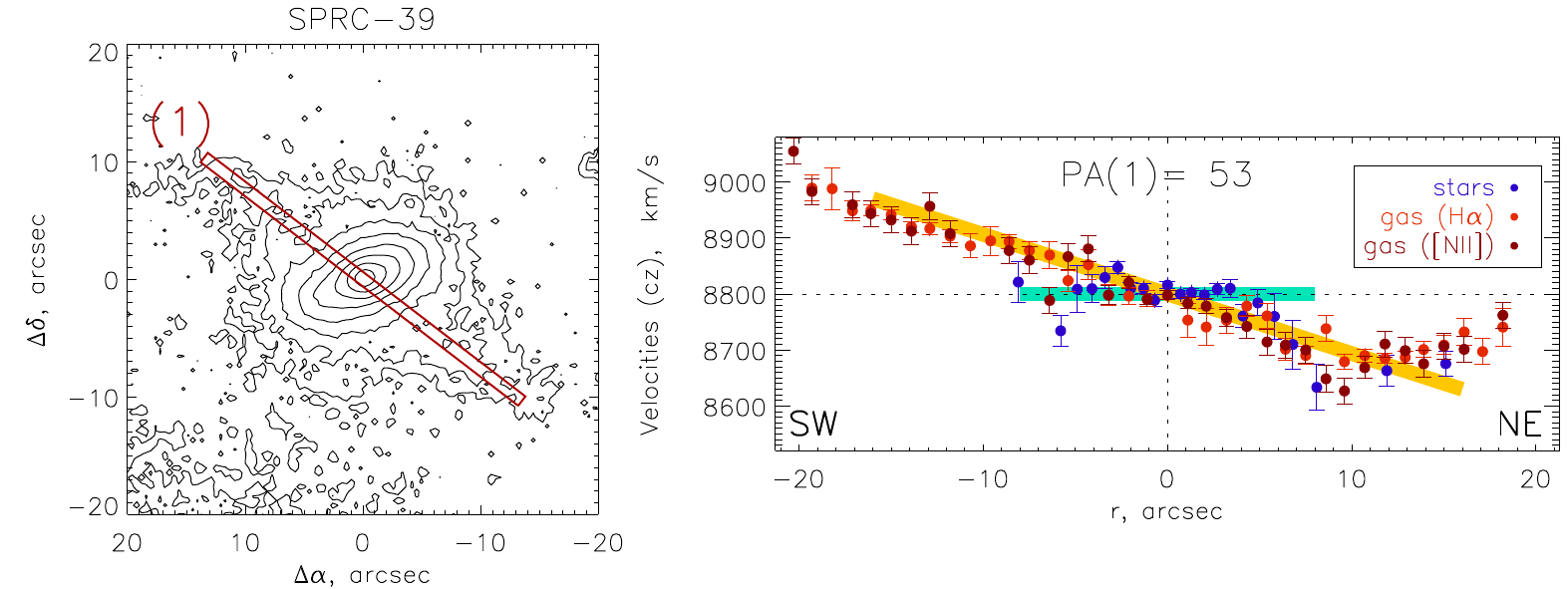}
\caption{Left: the  co-added image  in the $ugriz$ SDSS filters for the SPRC-10 , SPRC-14, and SPRC-39
galaxies.   The contours are in arbitrary units. The positions of the spectrograph slits are indicated.
Right: the distribution of line-of-sight velocities of gas and stars in
the corresponding sections. The dotted line marks the centre of
the galaxy and the value of systemic velocity. The thick coloured
lines schematically mark the kinematic components, rotating in
different planes: blue lines for the central galaxy and red lines for the possible polar ring.} \label{fig_spectr1}
\end{figure*}

\begin{figure*}
\includegraphics[width=0.97\textwidth]{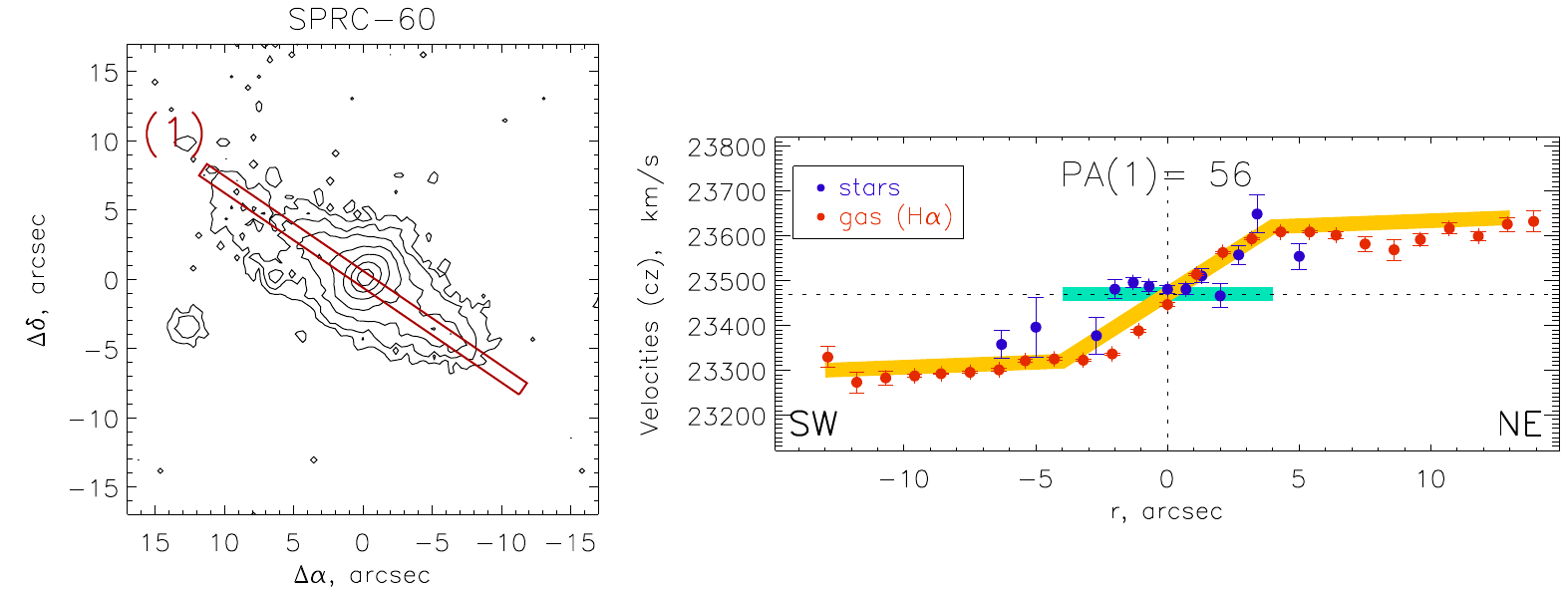}
\includegraphics[width=0.97\textwidth]{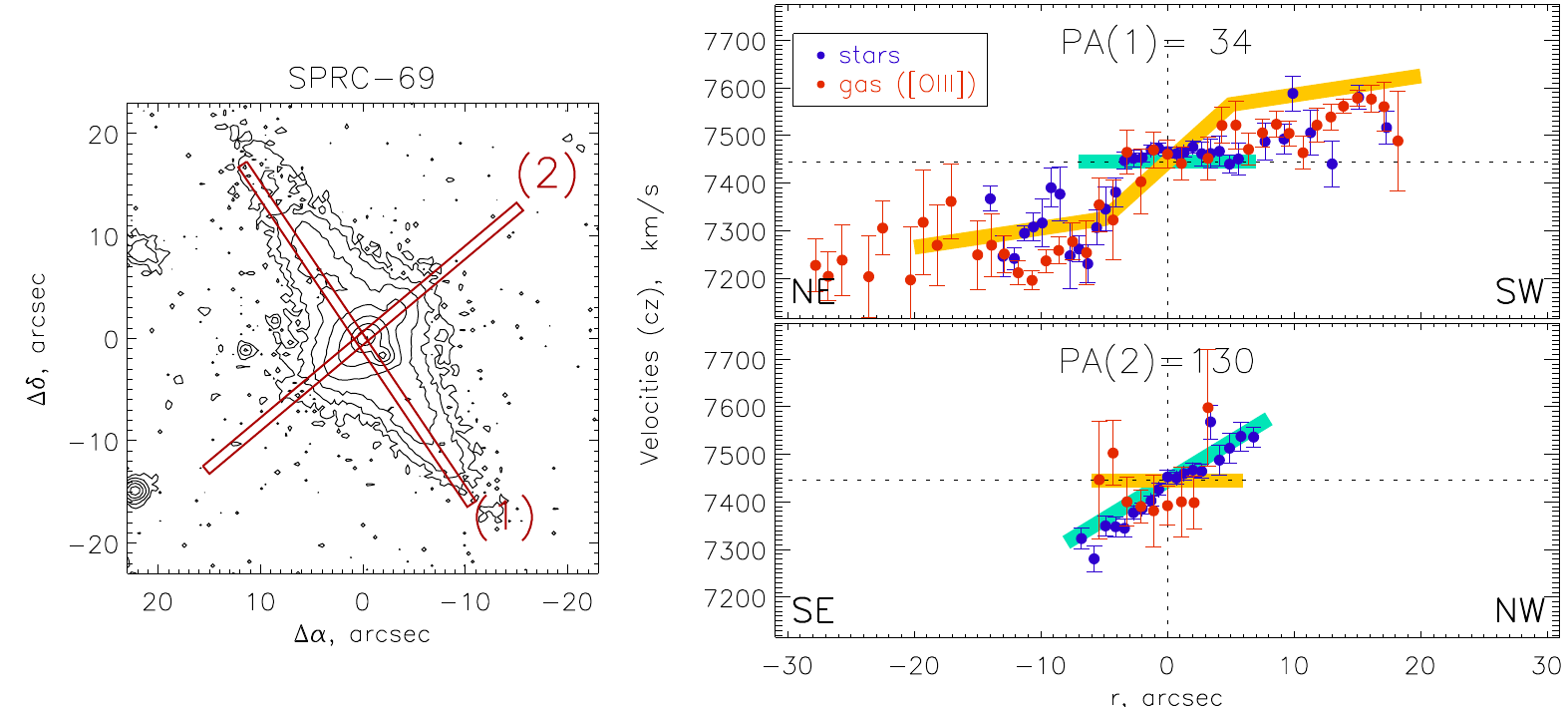}
\includegraphics[width=0.97\textwidth]{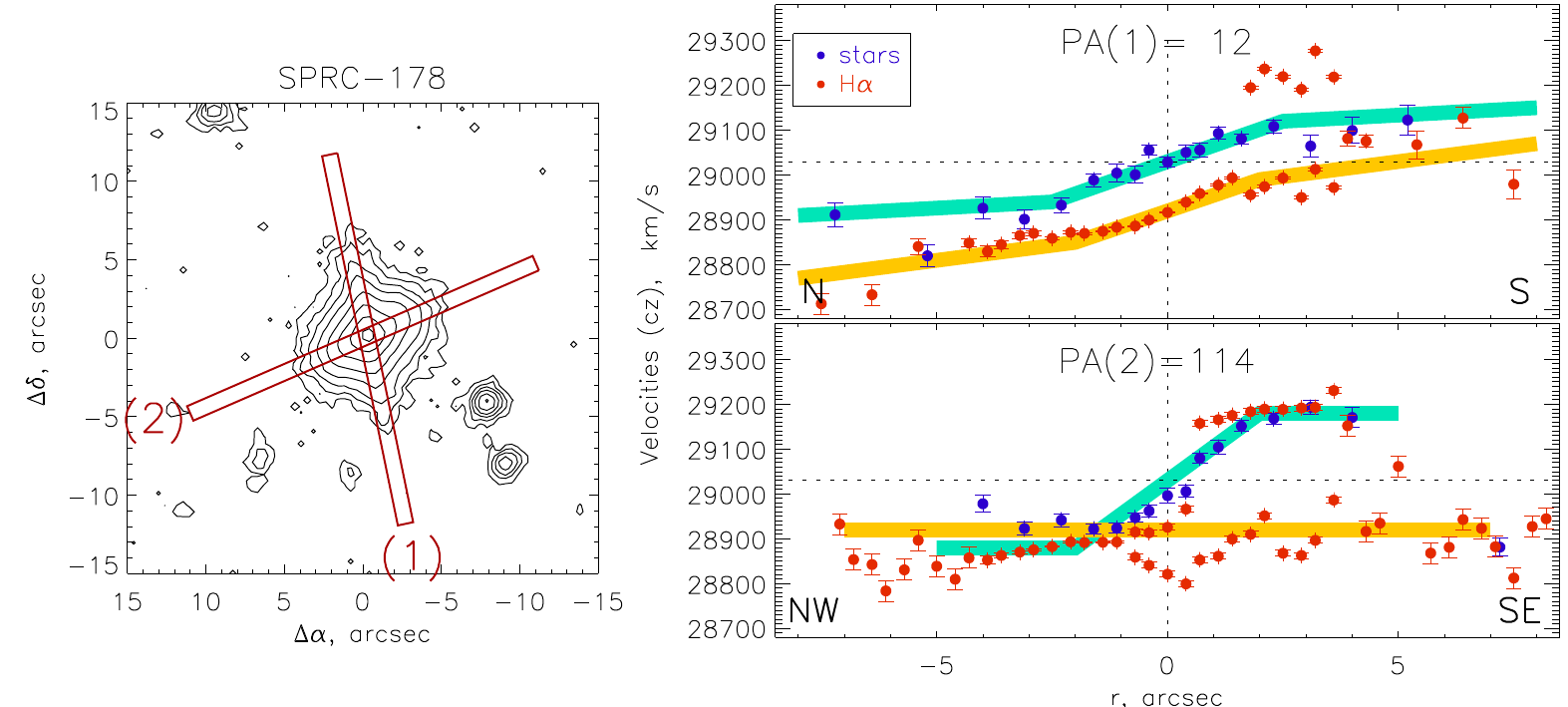}
 \contcaption{ The SPRC-60, SPRC-69, and SPRC-178 galaxies.}
\end{figure*}

\subsubsection{SPRC-10}

Two kinematic components are clearly detected here. The
distribution of line-of-sight velocities corresponds to that, expected
for a galaxy, in which all the ionized gas rotates in the plane,
strongly inclined to the inner stellar body. Along the first
section with $PA(1)=39\degr$ (the major axis of the ring)  a
significant gradient of line-of-sight velocities of ionized gas is
observed, while the line-of-sight velocity of stars in the inner region
$(r<3$ arcsec) barely varies. In the   slit position  along the major axis
of the central galaxy ($PA(2)=138\degr$) the situation is reversed
-- here we find a zero gradient of line-of-sight velocities of gas, and a
substantial gradient for the stellar component, where a
symmetrical rotation curve can clearly be traced. Interestingly,
in the   slit position  with $PA(1)=39\degr$ at $r>3-4$ arcsec on both
sides of the nucleus we can see a symmetrical increase in line-of-sight
velocities of stars, in the same direction with the ionized  gas in the ring. Apparently,
what we observe here are the stars of the polar ring.

\subsubsection{SPRC-14}

This galaxy also reveals two kinematic components, but the
observed pattern is more complex   than in the previous case. The scatter of line-of-sight velocities
of gas is maximal along the major axis of the ring
$PA(1)=35\degr$. Moreover, in the NE half, the ionized gas
emission extends almost twice farther than in the opposite direction away
from the centre. This  slit position almost exactly coincides  with the minor axis of the central disc, but line-of-sight velocities of stars are constant only at $r<4$ arcsec, while at
large distances they almost coincide with the velocities of gas in
the ring, except for the outermost region in the SW part of the
section. It means that just like in SPRC-10,  we observe the motion
of stars, belonging to the ring at a distance from the centre. At
the same time, line-of-sight velocities of ionized gas at $r=4-12$ arcsec
to the NE from the centre almost coincide with $V_{sys}$. This
behaviour can be explained by the fact that a part of the ionized
gas is located not only in the ring, but also rotates in the
central disc of the galaxy. Collisions with gas clouds of the ring
greatly entangle the observed picture. Velocities of gas
along the minor axis of the ring ($PA(2)=130\degr$) noticeably
differ from the systemic velocity, and near the nucleus we
can even observe the counter-rotation relative to the outer
region. This most likely indicates that the polar ring is
significantly curved.

\subsubsection{SPRC-39}

Here we managed to obtain only one section -- along the major axis
of the ring. Nonetheless, the observed picture is very similar to
what we have seen in SPRC-10  and SPRC-14. Namely, the line-of-sight
velocities of stars barely vary in the centre of ($r<6-8$ arcsec),
since the slit passes near the minor axis of the central disc. At
large distances, we can see the movement of stars in the ring. We
note here a strong asymmetry of the rotation curves of gas in both
directions from the centre -- in the SW part it continues to grow,
while to the NE from the centre it achieves plateau.

\subsubsection{SPRC-60}

There is only one section here as well, revealing a picture close
to that, described above for SPRC-39. In the central part, the
gradient of line-of-sight velocities of stars is nearly zero. We can see
the stars, belonging to the ring at increasing distances from the
nucleus. Unlike in SPRC-39, the rotation curve of gas in the ring
is symmetrical here on both sides from the centre.

\subsubsection{SPRC-69=II~Zw~92}

This is the most nearby galaxy we observed. The notes to the
catalogue by \citet{Zwicky1971} indicate that this galaxy ``appears
like Saturn''. The SDSS colour image clearly demonstrates how the
outer highly inclined ring surrounds the central galaxy
(Fig.~\ref{fig_example}). The results of spectral observations
confirm the presence of the polar ring, which reveals both ionized
gas and stars (see the   slit position with $PA(1)=34\degr$). The curve
of line-of-sight velocities of gas along the major axis of the ring is
asymmetrical and considerably more extended in the NE direction
from the nucleus. This may be caused by the peculiarities of
distribution of HII regions in the ring. Along the major axis of
the central body $PA(2)=130\degr$ we observe a linear increase in
the rotation velocity of stars, while the ionized gas is visible
only in the very centre.

\subsubsection{SPRC-178}

Judging from Fig.~\ref{fig_spectr1}, we observe here two galaxies
with similar line-of-sight velocities. Stellar absorptions are noticeable
in the more inclined  disc with a
systemic velocity of $V_{sys}=29030\km$ and the major axis with
$PA(2)=114\degr$. The gradient of  velocities of stars is
maximal along this direction, however, along $PA(1)=12\degr$ the
gradient is also nonzero, since the slit does not coincide with
the minor axis of the disc. The lines of ionized gas are brighter
in the second galaxy, seen nearly edge-on, with
$V_{sys}=28920\km$. Its rotation curve is clearly visible in the
section along the major axis with $PA(1)=12\degr$. The lines of
ionized gas in the SE part of the system noticeably bifurcate, we
can see here the rotation of the two galaxies separately along the
line-of-sight, which is particularly noticeable along
$PA(2)=114\degr$. In the NW part, the velocities of both discs in
the projection to the line-of-sight  almost coincide, and can
not be separated. Note that the section with $PA(2)=114\degr$ is
close to the minor axis of the galaxy, visible edge-on. Meanwhile,
the ionized gas with the corresponding systemic velocity is traced
relatively far from the galactic plane, up to $8$ arcsec (14 kpc)
in the SW direction. It is possible that what  we observe here is a
common gas envelope or a tidal flow, in which case we are dealing
with a pair of interacting galaxies.

\section{Discussion}
\label{sec_conclusion}

As a result of first spectral observations of galaxies from our
catalogue we were able to confirm the presence of polar rings in
five of them. The sixth object, SPRC-178, turned out to be a pair
of galaxies. All the confirmed candidates reveal both ionized gas
and stars in their polar orbits. We hope that a more in-depth
analysis of these and subsequent observations will allow us to say
more about the stellar population of   polar rings: their age,
metallicity and kinematics. Note that by now we  only knew of
three galaxies, in which the motion of stars in polar rings was
studied -- NGC~4650A \citep{SwatersRubin2003}, UGC~5119
\citep{Merkulova2008}, and UGC~2748 \citep{Merkulova2009}.

Along with the available literature data, the kinematic
confirmation of the presence of an outer polar component exists
for 10 galaxies from our catalogue. This is already a significant
supplement to the number of known PRGs.

\begin{figure}
\includegraphics[width=0.45\textwidth]{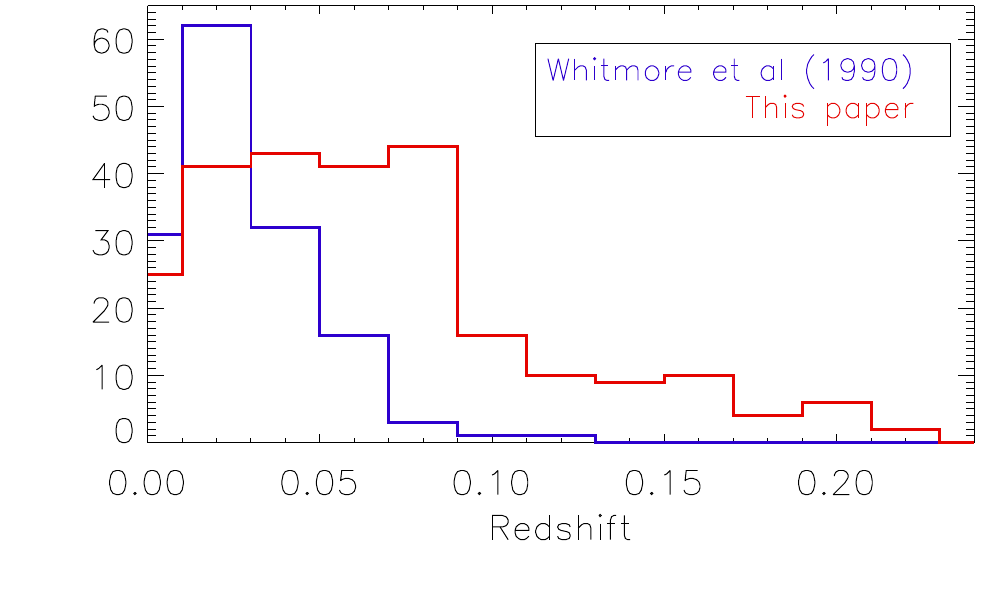}
\caption{Histograms of the distribution of PRG candidates by
redshift for the \citet{Whitmore1990} catalogue and the new list.}
\label{fig_z}
\end{figure}

Supplementing the PRC galaxies with the objects from our SPRC
catalogue, we have by 3 times increased the number of known
objects that are in one way or another connected with the
phenomenon of polar rings. Thereby, the number of ``genuine'' PRG
candidates has increased almost threefold, since the 33 PRC-A and
PRC-B objects were subjoined by   70 best candidates  from the SPRC.
Fig.~\ref{fig_z} compares the redshift distributions of galaxies
from \citet{Whitmore1990} and SPRC. It is evident that the new
catalogue contains a greater number of distant objects, starting
from $z>0.05$. A sharp decline in the number of candidates at
$z>0.1$ is obviously connected with the selection -- the images
obtained with $seeing\approx1$ arcsec allow to distinguish
structural features only in   galaxies having relatively large angular sizes. At the same
time, dozens of new nearby ($z<0.03$) PRG candidates were
discovered.

Therefore, even among the nearby galaxies, the PRGs occur several times
more frequently than it was previously thought. By a rough
estimate of \citet{Whitmore1990}, only 0.5 per cent of nearby S0 galaxies
possess polar rings, while this estimate could be significantly increased
in view of the rings, seen face-on. More accurate recent
estimates by \citet{Resh2011}, based on the constructed luminosity
function, prove that the rate of PRG occurrence among the nearby
galaxies   in the range of $M_B=-17^m...-22^m$ amounts to 0.13 per cent,
and adjusted for the effect of projection it reaches  0.4 per cent.
Employment of new catalogue objects can significantly increase
this value, since, according to the criteria, described in
Section \ref{sect_zoo}, the actual number of PRGs among all the
Galaxy Zoo galaxies is about 3 times greater than in our
catalogue. Moreover, the PRC was based on viewing the photographic
plates with a smaller limit on surface brightness, than the SDSS
data have. This is why the objects with fainter outer structures
made it into the SPRC catalogue. However, before reviewing the
luminosity function, we have to make the kinematic confirmation of
new candidates.

The galaxies with massive outer polar rings are of particular
interest. Formation of these rings can not always be
explained within the classical scenario of accretion or merging.
The best  example known to date is a nearby galaxy NGC~4650A,
possessing a massive self-gravitating stellar-gaseous ring. A recent
paper by \citet{Spavone2010} hypothesizes that such a polar disc
in this galaxy was formed by accretion of gas from the
intergalactic medium filaments. Among the objects of our catalogue
there are several galaxies with bright and extended polar rings:
SPRC-7, SPRC-27, and SPRC-40. \citet{Brosch2010} have shown that
the former galaxy can largely be regarded as a giant analogue of
NGC~4650A. Further detailed study of similar systems will help to
better understand the processes of gas accretion  during the
galaxy formation on the cosmological scales.

\section*{ACKNOWLEDGMENTS}
We have made use of the NASA/IPAC Extragalactic Database (NED)
which is operated by the Jet Propulsion Laboratory, California
Institute of Technology, under the contract with the National
Aeronautics and Space Administration.  This work was supported by
the Russian Foundation for Basic Research (project
no.~09-02-00870). AVM is also grateful to the Dynasty Foundation.
VPR acknowledges support from the ``Bourse de la Ville de Paris''.
We wish to thank  Victor Afanasiev for his great contribution to spectroscopy at the 6-m telescope,
Timur Fatkhullin for the help in observations in   August 2010,
 Ido Finkelman, who provided us with a list of new polar-ring candidates before their publication;
and  an anonymous referee for constructive advice, which   has helped us to improve the paper.

Funding for the SDSS has been provided by the Alfred P. Sloan
Foundation, the Participating Institutions, the National Science
Foundation, the U.S. Department of Energy, the National
Aeronautics and Space Administration, the Japanese Monbukagakusho,
the Max Planck Society, and the Higher Education Funding Council
for England. The SDSS Web Site is http://www.sdss.org/.

The SDSS is managed by the Astrophysical Research Consortium (ARC)
for the Participating Institutions. The Participating Institutions
are The University of Chicago, Fermilab, the Institute for
Advanced Study, the Japan Participation Group, The Johns Hopkins
University, Los Alamos National Laboratory, the
Max-Planck-Institute for Astronomy (MPIA), the
Max-Planck-Institute for Astrophysics (MPA), New Mexico State
University, University of Pittsburgh, Princeton University, the
United States Naval Observatory, and the University of Washington.

Many thanks also to the Galaxy Zoo (http://www.galaxyzoo.org) organizers and participants, especially
the  ring galaxies forum group. Galaxy Zoo has been supported in part by a Jim Gray research grant from
Microsoft, and by a grant from The Leverhulme Trust.

\newpage

\begin{table*}
\caption{Catalogue of PRG candidates} \label{tab_ABCD}
\begin{tabular}{rrrlrl}
\hline
\hline
\multicolumn{5}{c}{Best candidates}\\
SPRC&   R.A. (2000.0)& Dec (2000.0)& r (mag)&   cz, $\km$ & catalogue names      \\
   1 &  00 09 11.58 & -00 36 54.7 &  15.59 &  21960 &                  \\
   2 &  00 26 41.51 & +25 51 01.7 &  14.37 &      - &                  \\
   3 &  00 41 03.44 & -09 56 28.1 &  16.11 &  11063 &                  \\
   4 &  02 42 58.43 & -00 57 09.2 &  16.96 &  12888 &                  \\
   5 &  03 44 26.15 & -05 42 09.6 &  16.93 &   8364 &                  \\
   6 &  07 50 41.63 & +50 57 40.1 &  16.94 &   5586 &                  \\
   7 &  07 52 34.32 & +29 20 49.7 &  16.95 &  18032 &                  \\
   8 &  07 55 39.63 & +18 35 27.3 &  17.10 &      - &                  \\
   9 &  08 12 57.83 & +22 25 30.0 &  17.18 &  43402 &                  \\
  10 &  08 20 38.19 & +15 36 59.8 &  16.29 &  12736 &                  \\
  11 &  08 48 32.01 & +32 20 12.5 &  15.68 &  19733 &                  \\
  12 &  09 01 50.14 & +52 12 22.6 &  17.16 &        &                  \\
  13 &  09 14 53.66 & +49 38 24.0 &  15.37 &   9521 &                  \\
  14 &  09 18 15.97 & +20 22 05.3 &  14.73 &   9548 &     CGCG 121-053 \\
  15 &  09 36 34.63 & +21 13 57.8 &  14.43 &  10281 &                  \\
  16 &  09 42 07.35 & +36 24 17.2 &  16.95 &  18028 &                  \\
  17 &  09 59 11.85 & +16 28 41.5 &  14.92 &   7914 &                  \\
  18 &  10 08 42.82 & +14 22 02.1 &  17.25 &  24538 &                  \\
  19 &  10 17 17.50 & +41 05 57.7 &  17.39 &  31524 &                  \\
  20 &  10 26 00.41 & +04 14 33.0 &  17.84 &  22287 &                  \\
  21 &  10 28 58.56 & +31 33 42.2 &  17.15 &  24353 &                  \\
  22 &  10 41 55.66 & +07 45 13.9 &  17.02 &  47932 &                  \\
  23 &  10 46 23.65 & +06 37 10.1 &  15.64 &   8330 &                  \\
  24 &  11 16 25.11 & +56 50 17.0 &  14.98 &  14133 &                  \\
  25 &  11 24 53.65 & +21 53 48.9 &  17.16 &  21836 &                  \\
  26 &  11 40 05.50 & +06 04 43.7 &  18.70 &      - &                  \\
  27 &  11 44 44.02 & +23 09 45.0 &  16.28 &  14508 &                  \\
  28 &  11 52 28.30 & +05 00 44.7 &  16.80 &  23188 &                  \\
  29 &  11 53 33.56 & +47 19 07.3 &  15.25 &  14208 &                  \\
  30 &  12 07 12.69 & +07 39 44.5 &  17.48 &  22527 &                  \\
  31 &  12 17 11.51 & +31 30 37.8 &  15.08 &  14913 &                  \\
  32 &  12 19 24.16 & +64 37 54.0 &  16.65 &  10480 &                  \\
  33 &  12 19 30.57 & +14 52 39.5 &  11.24 &   1358 &        NGC~4262  \\
  34 &  12 25 59.80 & +15 42 08.5 &  17.43 &  24375 &                  \\
  35 &  12 44 14.99 & +17 00 49.2 &  17.23 &  20256 &                  \\
  36 &  12 52 17.80 & +12 06 30.3 &  17.71 &        &                  \\
  37 &  12 53 09.76 & +49 48 32.5 &  16.38 &  20297 &                  \\
  38 &  13 08 14.92 & +28 44 07.5 &  15.90 &  11688 &                  \\
  39 &  13 08 16.92 & +45 22 35.2 &  16.01 &   8792 &                  \\
  40 &  13 11 31.46 & +36 16 54.6 &  17.38 &   1126 &        NGC~5014  \\
  41 &  13 25 33.22 & +27 22 46.7 &  15.88 &  18320 &                  \\
  42 &  13 39 04.59 & +02 09 49.5 &  15.04 &   7041 &        UGC~08634 \\
  43 &  13 41 34.83 & +37 26 25.7 &  17.67 &  51352 &                  \\
  44 &  13 45 06.11 & +27 30 14.3 &  16.38 &  34003 &                  \\
  45 &  13 52 39.38 & +01 06 09.1 &  16.05 &  21461 &                  \\
  46 &  13 55 00.36 & +24 10 05.7 &  17.68 &  38517 &                  \\
  47 &  13 59 41.70 & +25 00 46.1 &  14.49 &   9370 &                  \\
  48 &  14 14 20.82 & +27 28 04.4 &  15.04 &  16788 &                  \\
  49 &  14 14 46.25 & +30 25 17.8 &  15.98 &  20499 &                  \\
  50 &  14 15 59.68 & +19 24 30.6 &  16.89 &  23293 &                  \\
  51 &  14 18 16.05 & +20 31 34.3 &  17.17 &  22616 &                  \\
  52 &  14 18 25.60 & +25 30 06.7 &  15.07 &   4450 &     KUG 1416+257 \\
  53 &  14 19 18.45 & +01 50 49.1 &  17.73 &  24837 &                  \\
  54 &  14 39 48.62 & +21 46 20.3 &  15.85 &  11805 &                  \\
  55 &  14 43 48.70 & +16 09 29.7 &  16.98 &  25782 &                  \\
  56 &  15 11 14.09 & +37 02 37.7 &  14.81 &  16499 &   MCG +06-33-026 \\
  57 &  15 11 53.36 & +27 43 46.7 &  16.16 &  21093 &                  \\
  58 &  15 38 01.95 & +56 42 21.8 &  15.66 &      - &                  \\
  59 &  15 38 45.54 & +65 27 38.3 &  16.65 &      - &                  \\
  60 &  15 47 24.32 & +38 55 50.4 &  17.29 &  23519 &                  \\
 \hline
\end{tabular}
\end{table*}

  \begin{table*}
 \contcaption{ }
 \begin{tabular}{rrrlrl}
 \hline
SPRC &  R.A. (2000.0)   &  Dec (2000.0)  &  r (mag)     & cz, $\km$& catalogue names  \\
 \hline
  61 &  15 49 54.81 & +09 49 43.1 &  14.81 &  13753 &                  \\
  62 &  15 51 26.66 & +35 10 45.9 &  17.48 &      - &                  \\
  63 &  15 51 28.30 & +04 30 11.6 &  17.38 &  22184 &                  \\
  64 &  15 58 29.79 & +57 05 51.5 &  17.60 &      - &                  \\
  65 &  16 16 00.16 & +24 21 05.6 &  16.28 &  20038 &                  \\
  66 &  16 31 58.73 & +48 17 22.6 &  15.91 &  26207 &                  \\
  67 &  17 17 44.13 & +40 41 52.0 &  14.25 &   8325 &     CGCG 225-097 \\
  68 &  17 50 57.21 & +51 25 19.2 &  16.46 &      - &                  \\
  69 &  20 48 05.67 & +00 04 07.8 &  15.36 &   7396 &        II Zw 092 \\
  70 &  23 42 33.22 & +13 33 25.6 &  17.39 &  20595 &                  \\
\multicolumn{5}{c}{ }\\
\multicolumn{5}{c}{Good candidates}\\
  71 &  00 05 20.62 & +14 34 11.1 &  17.23 &  12459 &                  \\
  72 &  00 13 03.07 & +14 24 35.9 &  13.85 &   2031 &       UGC 0119   \\
  73 &  00 32 09.80 & +01 08 36.6 &  15.80 &  17714 &                  \\
  74 &  00 48 12.18 & -00 12 55.6 &  15.87 &  16925 &                  \\
  75 &  01 07 00.15 & +28 45 33.8 &  15.41 &      - &                  \\
  76 &  01 21 29.32 & +00 37 28.7 &  16.83 &  13058 &                  \\
  77 &  01 58 58.39 & -00 29 23.3 &  17.47 &  24318 &                  \\
  78 &  07 32 08.78 & +27 27 41.2 &  17.36 &      - &                  \\
  79 &  07 33 21.06 & +20 21 12.4 &  16.94 &      - &                  \\
  80 &  07 47 15.86 & +43 35 04.7 &  16.70 &  32568 &                  \\
  81 &  07 48 20.90 & +36 04 34.1 &  16.03 &  41043 &                  \\
  82 &  07 52 41.21 & +19 43 20.2 &  16.91 &  27861 &                  \\
  83 &  08 00 23.93 & +17 31 26.9 &  15.76 &   2061 &                  \\
  84 &  08 11 17.51 & +29 30 32.2 &  13.91 &   5970 &      CGCG 149-001\\
  85 &  08 14 07.20 & +25 14 48.5 &    16. &   7445 &    KUG 0811+253  \\
  86 &  08 23 23.98 & +30 53 18.3 &  17.45 &  48506 &                  \\
  87 &  08 42 00.51 & +13 12 29.7 &  17.48 &  36845 &                  \\
  88 &  08 56 04.40 & +34 42 34.9 &  17.68 &  26498 &                  \\
  89 &  09 07 56.24 & +28 26 33.7 &  17.83 &  50245 &                  \\
  90 &  09 43 02.33 & -00 48 50.0 &  15.94 &  20270 &                  \\
  91 &  09 57 31.05 & +35 33 22.8 &  17.46 &  63206 &                  \\
  92 &  09 58 44.65 & +12 59 20.9 &  16.75 &  57802 &                  \\
  93 &  09 59 05.51 & +02 38 09.9 &  15.02 &  23743 &                  \\
  94 &  09 59 39.90 & +11 24 22.3 &  17.81 &  23908 &                  \\
  95 &  10 05 57.71 & +64 40 54.6 &  17.53 &  17211 &                  \\
  96 &  10 06 13.91 & +63 13 21.9 &  17.20 &  34875 &                  \\
  97 &  10 17 30.50 & +21 21 01.9 &  16.96 &   6291 &                  \\
  98 &  10 18 00.74 & +01 21 16.3 &  17.28 &  37897 &                  \\
  99 &  10 34 35.22 & +44 46 23.0 &  17.40 &  50688 &                  \\
 100 &  10 35 42.04 & +26 07 33.0 &  14.39 &   1391 &    CGCG 124-041  \\
 101 &  10 36 02.78 & +13 10 27.7 &  17.55 &      - &                  \\
 102 &  10 41 09.76 & +06 49 12.2 &  14.53 &   9998 &      VIII Zw 102 \\
 103 &  10 43 31.49 & +32 15 05.1 &  17.35 &  24904 &                  \\
 104 &  10 45 22.03 & +55 57 40.0 &  14.05 &    944 &        NGC~3353  \\
 105 &  10 48 44.73 & +38 46 36.2 &  16.63 &   7462 &                  \\
 106 &  10 52 22.48 & +18 07 11.1 &  16.80 &  39593 &                  \\
 107 &  10 54 24.16 & +44 30 02.0 &  16.89 &   2419 &                  \\
 108 &  10 55 00.95 & +46 30 48.5 &  16.60 &  16159 &                  \\
 109 &  11 08 54.49 & +02 40 38.6 &  14.69 &  10756 &     VIII Zw 126  \\
 110 &  11 09 29.00 & -00 00 08.2 &  16.62 &   7914 & 2dFGRS N368Z105  \\
 111 &  11 17 21.04 & +31 14 48.5 &  15.44 &   2126 &   KUG 1114+315B  \\
 112 &  11 19 41.61 & +41 01 40.4 &  17.64 &  21383 &                  \\
 113 &  11 35 13.68 & +65 01 41.7 &  18.35 &  44831 &                  \\
 114 &  11 35 14.67 & +59 49 09.5 &  17.41 &  33255 &                  \\
 115 &  11 43 46.46 & +33 42 16.7 &  17.34 &   9381 &                  \\
 116 &  11 45 12.95 & +35 05 10.4 &  16.36 &  20189 &                  \\
 117 &  11 51 36.04 & +16 39 51.4 &  15.93 &   4044 &    KUG 1149+169  \\
 118 &  11 52 07.90 & +38 01 57.6 &  17.49 &  61622 &                  \\
 119 &  11 52 44.10 & +55 43 53.5 &  16.36 &  18126 &                  \\
 120 &  12 03 46.23 & +10 27 43.7 &  17.13 &  20091 &                  \\
 \hline
\end{tabular}
\end{table*}

  \begin{table*}
 \contcaption{ }
 \begin{tabular}{rrrlrl}
 \hline
SPRC  &  R.A. (2000.0)   &  Dec (2000.0)  &  r (mag)     & cz, $\km$& catalogue names  \\
 \hline
 121 &  12 15 03.35 & +67 02 27.8 &  17.46 &  39752 &                  \\
 122 &  12 18 11.16 & +44 09 30.5 &  17.69 &   7446 & UGC~07340 NED01  \\
 123 &  12 46 56.39 & +51 36 49.3 &  13.74 &    501 &       UGC~07950  \\
 124 &  12 50 20.69 & +37 56 56.2 &  15.19 &  10531 &         IC 3828  \\
 125 &  12 57 44.58 & +02 41 30.2 &  19.33 &    926 &       UGC~08074  \\
 126 &  12 58 18.63 & +27 18 38.9 &  15.98 &   7446 &    KUG 1255+275  \\
 127 &  12 58 27.84 & +28 58 26.8 &  15.00 &   7139 &                  \\
 128 &  13 18 43.75 & +47 43 09.9 &  16.83 &  18555 &                  \\
 129 &  13 23 58.77 & -01 45 52.7 &  17.80 &  48476 &                  \\
 130 &  13 33 46.77 & +31 34 59.5 &  17.64 &  18679 &                  \\
 131 &  13 34 43.26 & +13 02 02.9 &  15.31 &  22073 &                  \\
 132 &  13 45 31.77 & +02 58 59.3 &  16.24 &  23292 &                  \\
 133 &  13 47 11.08 & +32 53 08.1 &  16.56 &  30776 &                  \\
 134 &  13 50 59.06 & +33 43 15.3 &  16.09 &  27981 &                  \\
 135 &  13 57 25.69 & +47 27 07.9 &  16.48 &  39584 &                  \\
 136 &  14 04 54.74 & +12 42 16.9 &  14.59 &   4183 &UGC~09002 = VV328 \\
 137 &  14 09 21.48 & +21 48 02.3 &  17.62 &  35306 &                  \\
 138 &  14 10 46.45 & +11 51 29.9 &  15.46 &  26663 &                  \\
 139 &  14 11 18.56 & +07 23 52.8 &  16.74 &  31072 &                  \\
 140 &  14 11 42.58 & +18 40 03.8 &  17.59 &  23911 &                  \\
 141 &  14 20 44.19 & +50 00 40.1 &  16.97 &  21141 &                  \\
 142 &  14 22 25.37 & +21 08 35.5 &  17.80 &  31076 &                  \\
 143 &  14 24 43.86 & +53 51 50.6 &  16.11 &      - &                  \\
 144 &  14 25 03.87 & +27 29 10.4 &  16.83 &  42238 &                  \\
 145 &  14 28 16.47 & +00 00 08.3 &  16.99 &  15915 &                  \\
 146 &  14 32 31.17 & +15 29 43.8 &  16.24 &  18161 &                  \\
 147 &  14 41 19.18 & +07 47 34.8 &  15.71 &   1770 &    KUG 1438+080  \\
 148 &  14 41 32.86 & +23 42 55.2 &  17.00 &  11202 &                  \\
 149 &  14 46 45.71 & +58 30 50.8 &  16.70 &  24013 &                  \\
 150 &  14 48 17.26 & +60 18 34.3 &  15.93 &  15205 &                  \\
 151 &  14 51 26.30 & +18 01 38.5 &  17.17 &     -  &                  \\
 152 &  14 53 16.45 & +16 32 45.1 &  16.77 &  13235 &                  \\
 153 &  14 54 02.24 & +36 17 41.4 &  17.89 &  55856 &                  \\
 154 &  14 57 10.71 & +27 56 30.6 &  16.67 &  17963 &                  \\
 155 &  15 06 39.12 & +27 00 24.2 &  16.82 &  17525 &                  \\
 156 &  15 16 21.58 & +08 06 04.7 &  15.07 &   9275 &      CGCG 049-084\\
 157 &  15 24 46.34 & +33 48 26.2 &  16.20 &  24675 &                  \\
 158 &  15 25 28.59 & +02 46 44.2 &  17.03 &  34733 &                  \\
 159 &  15 30 04.63 & +14 57 55.8 &  17.35 &  45142 &                  \\
 160 &  15 33 22.82 & +33 29 33.4 &  16.66 &  23050 &                  \\
 161 &  15 36 19.28 & +33 46 27.7 &  17.19 &  22364 &                  \\
 162 &  15 37 03.02 & +59 06 58.5 &  17.35 &  27065 &                  \\
 163 &  15 44 42.15 & +11 21 59.9 &  16.89 &  14333 &                  \\
 164 &  15 47 07.11 & +34 25 36.1 &  16.95 &   1438 &                  \\
 165 &  15 49 53.86 & +64 14 03.7 &  17.40 &      - &                  \\
 166 &  15 52 36.10 & +33 11 37.5 &  17.40 &  22519 &                  \\
 167 &  15 56 42.98 & +48 02 53.1 &  14.52 &   5492 &                  \\
 168 &  15 57 14.20 & +26 28 54.3 &  16.38 &  21630 &                  \\
 169 &  16 00 13.82 & +17 50 53.8 &  16.43 &   2049 &                  \\
 170 &  16 02 09.77 & +06 38 51.4 &  16.26 &  22026 &                  \\
 171 &  16 05 14.60 & +41 11 18.3 &  16.36 &   2123 &                  \\
 172 &  16 18 11.02 & +56 44 09.0 &  16.65 &      - &                  \\
 173 &  16 21 28.25 & +28 38 24.8 &  14.23 &    890 &       UGC~10351  \\
 174 &  16 22 02.63 & +48 01 38.4 &  16.52 &   2448 &                  \\
 175 &  16 27 30.74 & +49 06 00.8 &  17.12 &  57192 &                  \\
 176 &  16 32 02.04 & +43 11 49.0 &  16.32 &   9269 &                  \\
 177 &  16 33 47.53 & +22 48 35.6 &  17.39 &  83807 &                  \\
 178 &  16 36 52.00 & +34 15 04.3 &  17.14 &  28935 &                  \\
 179 &  17 03 08.79 & +46 22 25.1 &  13.66 &   9610 &    CGCG 252-017  \\
 180 &  17 04 06.73 & +38 52 57.6 &  16.11 &  10604 &                  \\
 \hline
\end{tabular}
\end{table*}

  \begin{table*}
 \contcaption{ }
 \begin{tabular}{rrrlrl}
 \hline
SPRC  &  R.A. (2000.0)   &  Dec (2000.0)  &  r (mag)     & cz, $\km$& catalogue names  \\
 \hline
 181 &  17 08 17.84 & +29 53 40.0 &  17.04 &  44791 &                  \\
 182 &  19 33 27.34 & +35 49 30.1 &  15.26 &   4490 &                  \\
 183 &  21 37 48.61 & -07 40 28.1 &  16.10 &  17439 &                  \\
 184 &  22 08 21.70 & +18 27 15.1 &  14.80 &   1141 &      II Zw 166   \\
 185 &  23 12 32.10 & -00 06 37.0 &  14.86 &   8342 &                  \\
\multicolumn{5}{c}{ }\\
\multicolumn{5}{c}{Related objects}\\
 186 &  00 48 09.60 & -00 54 44.2 &  14.64 &  10182 &                  \\
 187 &  01 12 30.14 & -02 50 52.1 &  14.47 &  19241 &                  \\
 188 &  03 34 06.08 & +01 05 40.0 &  15.37 &  14235 &                  \\
 189 &  07 29 09.07 & +26 24 03.9 &  14.15 &   4957 &    CGCG 117-045  \\
 190 &  07 41 54.78 & +38 29 53.8 &  16.23 &   3513 &                  \\
 191 &  08 19 55.44 & +55 07 56.1 &  17.56 &  16523 &                  \\
 192 &  08 23 01.20 & +32 00 53.5 &  15.02 &  19246 &                  \\
 193 &  08 31 25.79 & +40 57 22.4 &  13.35 &   7309 &       UGC~04449  \\
 194 &  08 39 37.31 & +37 15 50.3 &  16.50 &  48491 &                  \\
 195 &  09 08 17.46 & +21 26 36.0 &  12.43 &   2720 &        NGC~2764  \\
 196 &  09 43 12.00 & +31 55 43.3 &  11.42 &   1566 &        NGC~2968  \\
 197 &  10 39 26.88 & +47 56 49.6 &  14.09 &    858 &       UGC~05791  \\
 198 &  10 41 09.76 & +06 49 12.2 &  14.53 &   9998 &      VIII Zw 102 \\
 199 &  10 45 46.87 & -01 12 01.0 &  15.85 &  25410 &                  \\
 200 &  11 18 36.19 & +24 52 02.9 &  14.57 &   7503 &     CGCG 126-026 \\
 201 &  11 23 38.82 & +53 50 31.7 &  12.40 &   2889 &         NGC~3656 \\
 202 &  11 35 07.51 & +29 53 27.7 &  14.43 &  13868 &        UGC~06557 \\
 203 &  11 37 53.79 & +21 58 52.3 &  13.05 &   8726 &         NGC~3753 \\
 204 &  11 54 43.08 & +33 32 12.4 &  14.73 &   9617 &        UGC~06882 \\
 205 &  11 55 38.34 & +43 02 45.1 &  13.29 &   7130 &        UGC~06901 \\
 206 &  12 00 04.42 & +58 36 21.0 &  14.75 &      - &   MCG +10-17-123 \\
 207 &  12 15 02.93 & +35 57 32.2 &  13.93 &    941 &        UGC~07257 \\
 208 &  12 16 02.45 & +06 04 18.0 &  17.34 &  30411 &                  \\
 209 &  12 40 30.88 & +05 52 10.6 &  16.05 &     -  &                  \\
 210 &  12 44 26.53 & -06 31 51.2 &  14.57 &  15483 &                  \\
 211 &  12 47 11.44 & +19 27 51.6 &  12.67 &   6789 &         NGC~4685 \\
 212 &  13 08 41.16 & +54 50 58.9 &  19.32 &      - &                  \\
 213 &  13 10 05.84 & +34 10 52.0 &  14.70 &    806 &                  \\
 214 &  13 35 37.37 & +13 19 32.5 &  14.46 &  12781 &        UGC~08576 \\
 215 &  13 41 17.08 & +26 16 19.4 &  15.10 &  19169 &                  \\
 216 &  13 44 50.67 & +55 43 05.9 &  15.95 &  35545 &                  \\
 217 &  14 01 27.22 & +31 38 47.0 &  15.25 &      - &                  \\
 218 &  14 32 39.84 & +36 18 08.0 &  12.43 &   3972 &         NGC~5675 \\
 219 &  14 41 32.45 & +46 40 34.6 &  16.12 &  25384 &                  \\
 220 &  15 12 02.25 & +21 17 53.4 &  14.07 &   4707 &        UGC~09763 \\
 221 &  15 17 25.86 & -00 08 05.5 &  16.99 &  15903 &                  \\
 222 &  15 17 49.18 & +04 09 45.4 &  13.29 &  11058 &  UGC~09804 NED01 \\
 223 &  15 20 17.21 & +14 04 31.5 &  17.53 &      - &                  \\
 224 &  16 12 34.06 & +28 19 08.2 &  14.53 &   9412 &    KUG 1610+284A \\
 225 &  16 19 23.92 & +17 38 43.6 &  15.25 &  20900 &                  \\
 226 &  16 20 01.60 & +37 35 18.1 &  16.79 &  20679 &                  \\
 227 &  16 28 20.56 & +32 48 34.8 &  14.93 &  10094 &         NGC~6161 \\
 228 &  16 33 46.26 & +49 50 08.4 &  14.11 &      - &   MCG +08-30-028 \\
 229 &  16 37 57.48 & +11 36 08.5 &  15.27 &  23512 &                  \\
 230 &  16 41 50.51 & +25 15 44.7 &  14.97 &  14212 &                  \\
 231 &  17 38 43.26 & +57 14 21.1 &  14.34 &   8800 &        UGC~10935 \\
 232 &  20 56 35.72 & -03 22 23.5 &  16.44 &   5842 &                  \\
 233 &  21 13 32.62 & -01 22 47.5 &  13.83 &  13669 &     CGCG 375-014 \\
 234 &  21 23 39.15 & -00 22 35.1 &  16.70 &  18582 &                  \\
 235 &  22 29 57.15 & +01 35 53.9 & 16.23: &      - &                  \\
 236 &  23 25 14.20 & +15 14 41.4 &  14.92 &  13025 &                  \\
 237 &  23 30 13.64 & +15 45 39.7 &  13.49 &   4212 &UGC~12633         \\
 238 &  23 39 57.61 & +00 08 06.7 &  15.44 &  18040 &     MCG-01-60-024\\
 \hline
\end{tabular}
\end{table*}

  \begin{table*}
 \contcaption{ }
 \begin{tabular}{rrrlrl}
 \hline
SPRC  &  R.A. (2000.0)   &  Dec (2000.0)  &  r (mag)     & cz, $\km$& catalogue names  \\
 \hline
 \multicolumn{5}{c}{Possible face-on rings }\\
 239 &  00 16 15.19 & -10 14 41.1 &  16.27 &  25566 &                  \\
 240 &  00 34 36.18 & -09 35 36.8 &  18.05 &      - &                  \\
 241 &  02 41 57.00 & -06 47 32.3 &  14.15 &   5267 &   MCG -01-07-035 \\
 242 &  07 35 50.48 & +32 35 06.3 &  17.68 &  60453 &                  \\
 243 &  07 53 25.87 & +15 06 56.8 &  17.08 &  30797 &                  \\
 244 &  07 58 48.20 & +38 21 50.5 &  17.86 &  29092 &                  \\
 245 &  08 05 05.38 & +40 10 28.4 &  14.24 &  15024 &     CGCG 207-027 \\
 246 &  08 08 21.30 & +07 17 12.6 &  16.99 &      - &                  \\
 247 &  08 12 12.08 & +44 39 20.5 &  17.63 &  55134 &                  \\
 248 &  08 24 01.83 & +16 26 38.2 &  17.62 &  45769 &                  \\
 249 &  08 47 08.70 & +19 37 51.7 &  14.03 &   9438 &       UGC~04596  \\
 250 &  08 47 41.69 & +13 25 08.8 &  13.46 &   2071 &     UGC 04599    \\
 251 &  08 49 01.21 & +41 11 57.2 &  14.08 &   8824 &     UGC 04609    \\
 252 &  10 24 01.96 & +10 19 15.8 &  17.36 &  57101 &                  \\
 253 &  10 43 15.06 & +10 02 06.0 &  16.33 &  26880 &                  \\
 254 &  11 01 45.49 & +25 26 44.4 &  17.36 &  43410 &                  \\
 255 &  11 02 16.90 & +38 23 20.2 &  16.53 &  27444 &                  \\
 256 &  11 10 49.29 & +13 36 34.1 &  16.27 &  21033 &                  \\
 257 &  11 30 19.31 & +58 16 40.8 &  17.62 &  59052 &                  \\
 258 &  11 30 40.21 & +50 37 04.6 &  15.31 &  17384 &                  \\
 259 &  11 37 13.65 & +23 17 21.9 &  17.18 &  14815 &                  \\
 260 &  11 45 30.25 & +09 43 44.8 &  14.48 &   6399 &    CGCG 068-056  \\
 261 &  12 01 43.47 & +00 10 59.2 &  18.32 &  31300 &                  \\
 262 &  12 49 15.77 & +04 39 24.5 &  20.36 &   2666 &       UGC~07976  \\
 263 &  13 01 13.66 & +58 06 44.2 &  15.02 &   8444 &  MCG +10-19-014  \\
 264 &  13 16 14.69 & +41 29 40.1 &  14.24 &   6217 &        MRK 1477  \\
 265 &  13 20 22.67 & -00 24 07.6 &  17.82 &  32954 &                  \\
 266 &  13 20 54.29 & +41 35 35.5 &  17.09 &  46536 &                  \\
 267 &  13 37 39.21 & +54 35 24.8 &  16.15 &  20490 &                  \\
 268 &  13 40 04.76 & +64 03 02.0 &  17.45 &  45640 &                  \\
 269 &  14 04 53.72 & +12 43 18.1 &  14.25 &   4093 &UGC~09002 = VV328a\\
 270 &  14 09 31.41 & +50 08 17.3 &  15.66 &  21759 &                  \\
 271 &  14 51 48.17 & +16 47 27.7 &  16.06 &  35878 &                  \\
 272 &  15 45 47.04 & +14 06 13.5 &  15.96 &  22091 &                  \\
 273 &  16 06 07.97 & +10 34 06.0 &  17.16 &  51450 &                  \\
 274 &  16 33 09.59 & +34 55 34.7 &  14.80 &  10465 &    KUGX~1631+350 \\
 275 &  20 53 53.46 & -00 58 05.4 &  17.46 &  31807 &                  \\
\hline
\end{tabular}
\end{table*}

\label{lastpage}

\end{document}